\documentclass[11pt,a4paper,english]{article}

\usepackage[utf8]{inputenc}
\usepackage{amsmath,amssymb,amsthm,amsfonts}
\usepackage{color}
\usepackage{graphicx}
\usepackage{url}
\usepackage{xargs}
\usepackage{paralist}
\usepackage[export]{adjustbox}
\usepackage{lineno}
\usepackage{fullpage}
\usepackage{enumitem}
\usepackage[lined,tworuled,linesnumbered]{algorithm2e}
\usepackage{yhmath}

\usepackage{float}
\newfloat{algorithm}{t}{alg}

\usepackage[labelformat=simple]{subcaption} 
\usepackage[font=small,width=0.95\textwidth]{caption}
\usepackage[pdftex]{hyperref}
\usepackage{kvoptions-patch}
\usepackage{cleveref}

\DeclareMathAlphabet{\mathpzc}{OT1}{pzc}{m}{it}
\graphicspath{{fig/}{./fig/}}

\newcommand{\TheTitle}{Efficient computation of minimum-area rectilinear convex hull under rotation and generalizations}

\AtBeginDocument{%
  \maketitle
  \hypersetup{%
    pdftitle = {\TheTitle},
    pdfauthor = {Carlos Alegr\'{i}a-Galicia, David Orden, Carlos Seara, Jorge Urrutia}
  }%
  {\footnotetext{
    \begin{minipage}[l]{0.2\textwidth}
      \includegraphics[trim=10cm 6cm 10cm
      5cm,clip,scale=0.15]{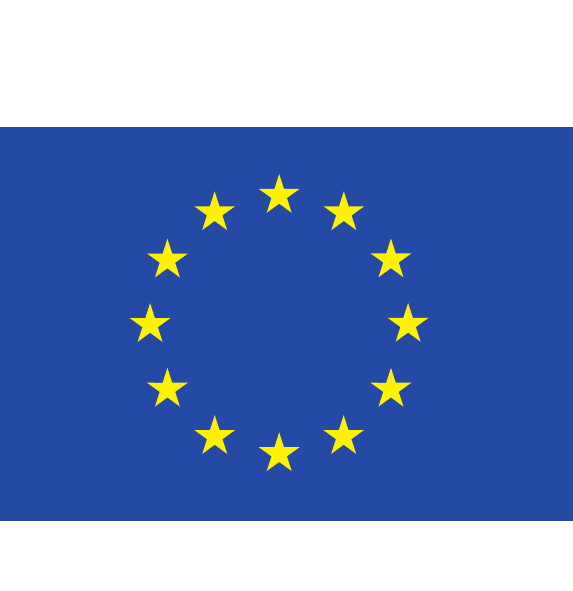}
    \end{minipage}
    \hspace{-1.5cm}
    \begin{minipage}[l][1cm]{0.8\textwidth}
      This project has received funding from the European Union's
      Horizon 2020 research and innovation programme under the Marie
      Sk\l{}odowska-Curie grant agreement No 734922.
    \end{minipage}%
  }}
}

\setlist[enumerate,1]{label=\textbf{\arabic*.},ref=\arabic*}
\setlist[enumerate,2]{label=(\alph*),ref=\arabic{enumi}(\alph*)}

\theoremstyle{plain}
\newtheorem{theorem}{Theorem}
\newtheorem{lemma}{Lemma}
\newtheorem{corollary}{Corollary}

\newtheorem{observation}{Observation}

\crefname{subsection}{subsection}{subsections}
\Crefname{subsection}{Subsection}{Subsections}

\newcommand{\area}{\operatorname{area}}
\newcommand{\adjustedaccent}[1]{%
  \mathchoice{}{}
    {\mbox{\raisebox{-.5ex}[0pt][0pt]{$\scriptstyle#1$}}}
    {\mbox{\raisebox{-.35ex}[0pt][0pt]{$\scriptscriptstyle#1$}}}
}

\newcommand{\idos}{[0,2\pi)}

\newcommandx*{\X}         [1][usedefault, 1=\theta]       {x_{#1}} 
\newcommandx*{\Y}         [1][usedefault, 1=\theta]       {y_{#1}} 
\newcommand{\W} {\mathcal{W}}
\newcommand{\Wt} {\mathcal{W}_{\theta}}
\newcommandx*{\rt}        [3][usedefault, 1=i, 2=\theta, 3=P]  {R^{#1}_{#2}({#3})} 
\newcommandx*{\rp}        [2][usedefault, 1=i, 2=P]  {R^{#1}({#2})} 

\newcommandx*{\Vt}        [2][usedefault, 1=\theta, 2=P]  {\mathcal{V}_{#1}({#2})} 
\newcommand{\V} {\mathcal{V}(P)}
\newcommandx*{\Tt}        [2][usedefault, 1=\theta, 2=P]  {\mathcal{T}_{#1}({#2})} 
\newcommandx*{\St}        [2][usedefault, 1=\theta, 2=P]  {\mathcal{S}_{#1}({#2})} 

\newcommandx*{\ch}        [1][usedefault, 1=P]            {\mathcal{CH}({#1})} 
\newcommandx*{\Ac}        [1][usedefault, 1=P]            {\mathcal{A}({#1})} 
\newcommandx*{\A}         [2][usedefault, 1=i, 2=P]       {A_{#1}({#2})} 
\newcommandx*{\link}      [2][usedefault, 1=i, 2=j]       {\ell_{{#1},{#2}}} 
\newcommandx*{\lpt}       [2][usedefault, 1=i, 2=j]       {p_{#2}^{#1}} 
\newcommandx*{\ldisk}     [2][usedefault, 1=i, 2=j]       {D_{{#1},{#2}}} 
\newcommandx*{\lrad}      [2][usedefault, 1=i, 2=j]       {r_{{#1},{#2}}} 
\newcommandx*{\larc}      [2][usedefault, 1=u, 2=v]       {\overset{\adjustedaccent{\smallfrown}}{\ell}_{{#1},{#2}}} 
\newcommandx*{\edisk}     [1][usedefault, 1=i]            {D_{#1}} 
\newcommandx*{\erad}      [1][usedefault, 1=i]            {r_{#1}} 
\newcommandx*{\polygon}   [1][usedefault, 1=\theta]       {\mathcal{P}(#1)}
\newcommandx*{\triangles} [2][usedefault, 1=\theta, 2=i]  {\triangle_{#2}(#1)}
\newcommandx*{\squares}   [2][usedefault, 1=\theta, 2=i]  {\square_{#2}(#1)}

\newcommandx*{\qd}   [3][usedefault, 1=p, 2=q, 3=\theta]  {Q_{#3}\left({#1},{#2}\right)}
\newcommandx*{\arc}  [2][usedefault, 1=a, 2=b] {\wideparen{{#1}{#2}}}

\newcommandx{\rch} [1][usedefault, 1=P] {\mathcal{RH}({#1})} 
\newcommandx{\rcht} [2][usedefault, 1=P,  2=\theta] {\mathcal{RH}_{#2}({#1})} 
\newcommandx{\os}  [1][usedefault, 1=P] {\mathcal{O}} 
\newcommandx{\ost}  [1][usedefault, 1=\theta] {\mathcal{O}_{#1}} 
\newcommandx{\oh}  [2][usedefault, 1=P, 2=\os] {{#2}\mathcal{H}({#1})} 
\newcommandx{\oht} [3][usedefault, 1=P, 2=\theta, 3=\os] {{#3}\mathcal{H}_{#2}({#1})} 

\title{\TheTitle
\thanks{Extended abstracts related to this work were presented at the XIV Spanish Meeting on Computational Geometry~\cite{alegria_2012} and the 30\textsuperscript{th} European Workshop on Computational Geometry (EuroCG)~\cite{alegria_2014}.}
}
\date{}

\author{
  Carlos Alegr\'{i}a-Galicia\thanks{Dipartimento di Ingegneria, Universit\`{a} Roma Tre, Italy {\tt carlos.alegria@uniroma3.it}. Research supported by MIUR Proj. ``AHeAD'' n\textsuperscript{o} 20174LF3T8.}
  \and
  David Orden\thanks{Departamento de F\'{\i}sica y Matem\'aticas, Universidad de Alcal\'a, Spain, {\tt david.orden@uah.es}. Research supported by project MTM2017-83750-P of the Spanish Ministry of Science (AEI/FEDER, UE) and project PID2019-104129GB-I00 of the Spanish Ministry of Science and Innovation.}
  \and
  Carlos Seara\thanks{Departament de Matem\`{a}tiques, Universitat Polit\`{e}cnica de Catalunya, Spain, {\tt carlos.seara@upc.edu}. Research supported by projects Gen.\ Cat.\ DGR 2017SGR1640, MINECO MTM2015-63791-R and project PID2019-104129GB-I00 of the Spanish Ministry of Science and Innovation.}
  \and
  Jorge Urrutia\thanks{Instituto de Matem\'{a}ticas, Universidad Nacional Aut\'{o}noma de M\'{e}xico, {\tt urrutia@matem.unam.mx}. Research supported in part by  SEP-CONACYT 80268, PAPPIIT IN102117 Programa de Apoyo a la Investigaci\'on e Innovaci\'on Tecnol\'ogica UNAM.}
  }

\begin{document}

\begin{abstract}
Let $P$ be a set of $n$ points in the plane. We compute the value of $\theta\in [0,2\pi)$ for which the rectilinear convex hull of $P$, denoted by $\rcht$, has minimum (or maximum) area in optimal $O(n\log n)$ time and $O(n)$ space, improving the previous $O(n^2)$ bound.
Let $\os$  be a set of~$k$ lines through the origin sorted by slope and let $\alpha_i$ be the sizes of the $2k$ angles defined by pairs of two consecutive lines, $i=1, \ldots , 2k$. Let $\Theta_{i}=\pi-\alpha_i$ and $\Theta=\min\{\Theta_i \colon i=1,\ldots,2k\}$. We obtain:
(1) Given a set $\os$ such that $\Theta\ge\frac{\pi}{2}$, we provide an algorithm to compute the $\os$-convex hull of~$P$ in optimal $O(n\log n)$ time and $O(n)$ space; If $\Theta < \frac{\pi}{2}$, the time and space complexities are $O(\frac{n}{\Theta}\log n)$ and $O(\frac{n}{\Theta})$ respectively.
(2)~Given a set $\os$ such that $\Theta\ge\frac{\pi}{2}$, we compute and maintain the boundary of the ${\os}_{\theta}$-convex hull of~$P$ for $\theta\in [0,2\pi)$ in $O(kn\log n)$ time and $O(kn)$ space, or if $\Theta < \frac{\pi}{2}$, in $O(k\frac{n}{\Theta}\log n)$ time and $O(k\frac{n}{\Theta})$ space.
(3)~Finally, given a set $\os$ such that $\Theta\ge\frac{\pi}{2}$, we compute, in $O(kn\log n)$ time and $O(kn)$ space, the angle $\theta\in [0,2\pi)$ such that the ${\os}_{\theta}$-convex hull of $P$ has minimum (or maximum) area over all $\theta\in [0,2\pi)$.
\end{abstract}


\section{Introduction}\label{sec:intro}

Restricted-orientation convexity is a generalization of traditional convexity that stems from the notion of \emph{restricted-orientation geometry}, where the geometric objects under study comply with restrictions related to a fixed set of orientations. Restricted-orientation geometry started with the work of G\"uting~\cite{guting_1983} in the early 1980s as a generalization of the study of orthogonal polygons, whose edges are parallel to the coordinate axes.

Under what is known as \emph{orthogonal convexity} or \emph{ortho-convexity}~\cite{montuno_1982,nicholl_1983,rawlins_1988} a set is called \emph{ortho-convex} if its intersection with any line parallel to a coordinate axis is either empty or connected. The \emph{rectilinear convex hull} of a geometric object~$S$, denoted as $\rch[S]$, is the smallest ortho-convex set that contains $S$, while Ottmann et al.~\cite{ottmann_1984} considered different definitions for the rectilinear convex hull~$\rch$ of a point set~$P$.
The rectilinear convex hull has been studied extensively since its introduction in the early 1980s, having been applied in several research fields, including illumination~\cite{abello_1998}, polyhedron reconstruction~\cite{biedl_2011}, geometric search~\cite{son_2014}, and VLSI circuit layout design~\cite{uchoa_1999}.

Researchers have also studied the rectilinear convex hull of colored point sets~\cite{alegria_2013}, and developed generalizations of orthogonal convexity~\cite{fink_2004,franek_2009,matousek_1998} along with related computational results~\cite{alegria_2014,alegria_2015_1,alegria_2018}. We use the definition of $\mathcal{O}$-convexity  introduced by Fink and Wood~\cite{fink_2004}: given the set $\mathcal{O}$ of lines through the origin, a set is  called $\mathcal{O}$-convex if its intersection with any line parallel to a line in $\mathcal{O}$ is either empty or connected. The $\mathcal{O}$-convex hull of a geometric object~$S$ is the smallest $\mathcal{O}$-convex set containing $S$; we will denote it as~$\oh[S]$. As in~\cite{ottmann_1984}, different definitions can be considered for the $\mathcal{O}$-convex hull of a point set~$P$, denoted as~$\oh$.

The $\mathcal{O}$-hull has applications in describing the shape of a point set in geometric information systems~\cite{bonerath_2019}, in distributed algorithms~\cite{daymude_2020}, and in the identification of Pareto-optimal solutions~\cite{Xu_pareto2015}. It also has applications when requiring a concave enclosing shape: In machine learning~\cite{lopez_2013}, the $\mathcal{O}$-hull allows us to extend support vector machines to use separating polygonal chains instead of separating lines. In manufacturing~\cite{griffiths_2019}, an optimal orientation is found by projecting the 3D object onto a 2D polygon and `digging' its convex hull edges. This is directly related to the problem solved in this paper; finding an optimal orientation to dig the convex hull of a point set with a given set of wedges. For potential applications, it is relevant to note that rotation-dependent and minimum area enclosing shapes are commonly used in form-shape analysis~\cite{biswas_2008,dutt_2011,jacob_2004,ZTYGJ_2010}, as well as in feature classification~\cite{lankveld_2011,sheikhi_2015}.

\subsection{Our work}

The $\theta$-rectilinear convex hull of a point set $P$, denoted as the $\rcht$-hull of $P$, is the $\mathcal{O}$-hull of a point set when $\mathcal{O}$ consists of two lines obtained by rotating the coordinate axes by an angle $\theta$.
The problem of finding the angle $\theta$ such that the area of $\rcht$ is minimized was introduced by Bae et al.~\cite{bae_2009}; they obtained an algorithm to find it in $O(n^2)$ time and $O(n)$ space.

In this paper we improve the results of Bae et al.~\cite{bae_2009} by providing an optimal $O(n \log n)$ time and $O(n)$ space algorithm to find an angle $\theta$ such that $\rcht$ is minimized.

Our algorithm maintains two structures. The first one maintains the set of vertices of~$\rcht$ as  $\theta$ increases from $0$ to $2\pi$. The second structure maintains the overlapping intervals, defined as the angular intervals contained in~$[0,2\pi)$, where opposite staircases of the boundary of $\rcht$ overlap each other (more details in Section~\ref{sec:urch:oet}); we show how to maintain these intervals in $O(n\log n)$ time and $O(n)$ space. This is one of the main contributions of our paper, included in Section~\ref{sec:rectilinear_hull}.
The maintenance of the set of vertices of~$\rcht$ as  $\theta$ increases from $0$ to $2\pi$ was first studied by D\'{\i}az-B\'{a}\~{n}ez et al.~\cite{fitting} in optimal $O(n\log n)$ time and $O(n)$ space, while designing an algorithm to fit an orthogonal chain to a point set in the plane.

In Section~\ref{sec:3} we study $\mathcal{O}$-convex sets. We design algorithms to compute the $\mathcal{O}$-convex hull of point sets $P$, denoted as $\mathcal{OH}(P)$. We also study the problem of maintaining the $\oht$-hull of a point set as we simultaneously rotate all the lines in $\mathcal{O}$ by an angle $\theta$ as $\theta$ changes in value from $0$ to $2\pi$.

Our algorithms are sensitive to two parameters; the cardinal $k\ge 2$ of a set  $\os=\{\ell_1,\dots,\ell_k\}$, and the sizes of the angles $\alpha_i$ of the $2k$ sectors defined by pairs of consecutive lines in $\mathcal{O}$ (sorted by slope), $i=1,\ldots,2k$. More specifically, let $\Theta_{i}=\pi-\alpha_i$. The second parameter will be $\Theta=\min\{\Theta_i\colon i=1,\ldots,2k\}$, which measures how ``\emph{well-distributed}''  the orientations of the $k$ lines of $\os$ are. We will distinguish whether $\Theta\ge \frac{\pi}{2}$ or $\Theta < \frac{\pi}{2}$.

\subsection{Our Contributions}

Our contributions can be summarized as follows:

\begin{itemize}

\item We improve the $O(n^2)$ time complexity from Bae et al.~\cite{bae_2009}, computing in optimal $O(n\log n)$ time and $O(n)$ space the value of $\theta\in [0,2\pi)$ for which $\rcht$ has minimum (or maximum) area, also returning $\rcht$ (Theorem~\ref{thm:unoriented_rch}).

\item Given a set $\os$ of~$k$ lines such that $\Theta\ge\frac{\pi}{2}$, we provide an algorithm to compute $\oh$ in optimal $O(n\log n)$ time and $O(n)$ space (Theorem~\ref{thm:oriented-oh}).

\item Given a set $\os$ of~$k$ lines such that $\Theta<\frac{\pi}{2}$, we provide an algorithm to compute $\oh$ in $O(\frac{n}{\Theta}\log n)$ time and $O(\frac{n}{\Theta})$ space (Theorem~\ref{thm:oriented-oh-general}).

\item We generalize to $\os$-convexity the problem in Bae et al.~\cite{bae_2009}. We show that for a set $\os$ of $k$ lines, computing and maintaining the boundary of $\oht$ during a complete rotation for $\theta\in [0,2\pi)$ can be done in $O(kn\log n)$ time and $O(kn)$ space (Theorem~\ref{thm:unoriented-oh}) for $\Theta\geq \frac{\pi}{2}$ or in $O(k\frac{n}{\Theta}\log n)$ time and $O(k\frac{n}{\Theta})$ space (Theorem~\ref{thm:unoriented-oh-general}) for $\Theta<\frac{\pi}{2}$.

\item As a consequence, when $\Theta\geq\frac{\pi}{2}$, computing an interval of $\theta$ such that the boundary of $\oht$ has the minimum number of staircases, has the minimum number of steps, is connected, or has the minimum number of connected components, can be done in $O(kn\log n)$ time and $O(kn)$ space (Corollary~\ref{coro:steps-components_1}). When $\Theta<\frac{\pi}{2}$ we can do it in $O(k\frac{n}{\Theta}\log n)$ time and $O(k\frac{n}{\Theta})$ space (Corollary~\ref{coro:steps-components_2}).

\item Given a set $\os$ consisting of two \emph{non-perpendicular} lines through the origin, we show that computing and maintaining the boundary of $\oht$ during a complete rotation for $\theta\in [0,2\pi)$ can be done in $O(\frac{n}{\Theta}\log n)$ time and $O(\frac{n}{\Theta})$ space, where $\Theta$ is the smallest angle of the sectors defined by the two lines (Corollary~\ref{coro:2directions}).

\item Given a set $\os$ of $k$ lines such that $\Theta\ge\frac{\pi}{2}$, we provide an algorithm to compute $\oht$ with minimum (or maximum) area over all $\theta\in [0,2\pi)$ in $O(kn\log n)$ time and $O(kn)$ space (Theorem~\ref{teorem_4_1}).
\end{itemize}

\section{Rectilinear hull of a point set}\label{sec:rectilinear_hull}

Let $P=\left\{p_{1},\dots,p_{n}\right\}$ be a set of $n$ points in the plane in general position. Let $\ch$ denote the convex hull of $P$ and let $V=\left\{ p_{1},\dots,p_{h}\right\}$ be the set of vertices on its boundary as we meet them when we traverse it in counterclockwise order, starting at one of its vertices, which we label $p_1$.
Further, let $E=\left\{e_{1},\dots,e_{h}\right\}$ be the set of edges of $\partial(\ch)$, where $e_{i}=\overline{p_{i}p_{i+1}}$ and the indices are taken modulo $h$.

\begin{figure}[ht]
  \begin{center}
  \includegraphics[width=0.4\textwidth]{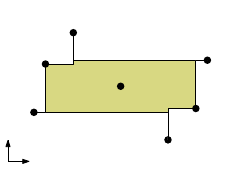}
  \qquad
  \includegraphics[width=0.4\textwidth]{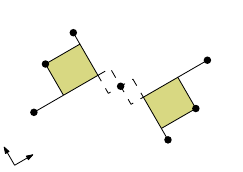}
  \end{center}
  \caption{Left: The rectilinear convex hull~$\rch$ of a point set~$P$. Right: The rectilinear convex hull~$\rcht$ of the same point set for $\theta=\pi/6$.}\label{fig:rcht}
\end{figure}

An \emph{open quadrant} is the intersection of two open half planes whose supporting lines are parallel to the $x$ and $y$ axes. An open quadrant is said to be \emph{$P$-free} if it contains no element of~$P$. The \emph{rectilinear convex hull} of $P$~\cite{ottmann_1984} is the set
\[
  \rch=\mathbb{R}^{2} \setminus\bigcup_{W \in \W}W,
\]
where $\W$ denotes the set of all $P$-free open quadrants.  See Figure~\ref{fig:rcht}, left, for an example. It is interesting to note that with this definition, the rectilinear convex hull could be disconnected~\cite{ottmann_1984}.

As in Bae et al.~\cite{bae_2009}, we will also consider the rectilinear convex hull when the axes are rotated by an angle~$\theta$, called the $\theta$-rectilinear convex hull of $P$:
\[
  \rcht=\mathbb{R}^{2} \setminus\bigcup_{W_{\theta} \in \Wt}W_{\theta},
\]
where $\Wt$ denotes the set of all $P$-free \emph{open $\theta$-quadrants}; i.e., the open quadrants obtained when the axes are rotated by~$\theta$. Figure~\ref{fig:rcht}, right, shows an example  where the $\pi/6$-rectilinear convex hull happens to be disconnected.

\subsection{Computing and maintaining \boldmath{$\rcht$}}\label{subsection:rcht}

In this subsection we describe an algorithm to compute and maintain $\rcht$ and other of its features as the axes are rotated by an angle~$\theta\in [0,2\pi)$. Our algorithm works in optimal $O(n\log n)$ time and $O(n)$ space, improving the $O(n^2)$ time complexity achieved by Bae et al.~\cite{bae_2009}.

To start, let us assume that $\theta=0$. Given two points $p_i,p_j\in P$ we say that \emph{$p_i$ dominates $p_j$}, denoted by $p_j\prec p_i$, if $x_j\le x_i$ and $y_j\le y_i$. It is well known that this defines a partial order in $P$. Note that this condition is equivalent to $p_j$ being in the closure of the open quadrant with apex $p_i$ which is a translation of the third quadrant. A point $p_i$ is called \emph{maximal} if there exists no $p_j\in P$ such that $i\neq j$ and $p_i\prec p_j$. The \emph{Set Maxima Problem}~\cite{preparata_1985} consists of finding all the maximal points of $P$ under this dominance~$\prec$. This problem can be solved optimally in $O(n\log n)$ time and $O(n)$ space~\cite{preparata_1985}.  Three additional partial orders can be obtained by rotating the plane by angles of $\pi/2$, $\pi$, and $3\pi/2$, and evidently for each of them, their set of maximal points can be also solved in $O(n \log n)$ time.

It is easy to see that the set of points in $P$ that belong to the boundary of its rectilinear convex hull is the union of the sets of maximal points in these four partial orders.

The set~$\Vt$ of points of~$P$ lying on the boundary of $\rcht$ will be called the set of \emph{vertices} of~$\rcht$. As above, for any fixed~$\theta$ the computation of $\rcht$ reduces to solving four set maxima problems, since
\begin{equation}
\label{eq:Vquadrants}
\Vt=V_\theta(P)\cup V_{\theta+\frac{\pi}{2}}(P)\cup V_{\theta+\pi}(P)\cup V_{\theta+\frac{3\pi}{4}}(P),
\end{equation}
for $V_\theta$ the set of maximal points of~$P$ with respect to the \emph{$\theta$-quadrant} defined by rotating the $x$ and $y$ axes by~$\theta$~\cite{bae_2009,fitting,ottmann_1984}. In order to keep track of the set~$\Vt$ as $\theta$ increases from $0$ to $2\pi$, we can use results from Avis et al.~\cite{theta-maxima_1999} and D\'{\i}az-B\'a\~nez et al.~\cite{fitting} as follows.

\begin{figure}[ht]
\begin{center}
\subcaptionbox{\label{fig:prel:interval_ini}}
{\includegraphics[width=0.32\textwidth]{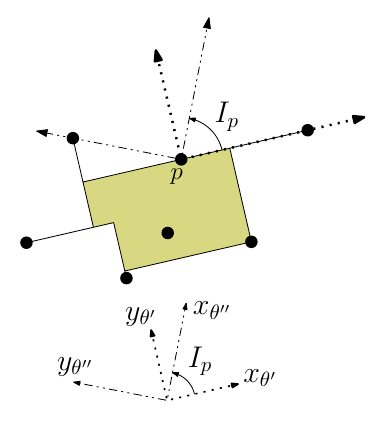}}
\subcaptionbox{\label{fig:prel:interval_middle}}
{\includegraphics[width=0.32\textwidth]{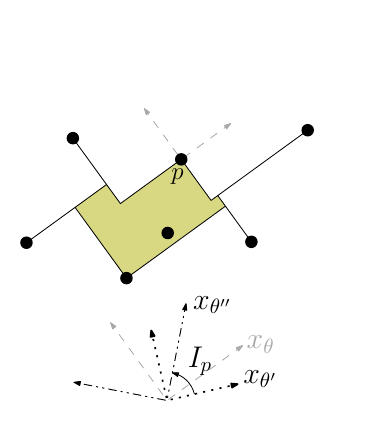}}
\subcaptionbox{\label{fig:prel:no_intervalo}}
{\includegraphics[width=0.32\textwidth]{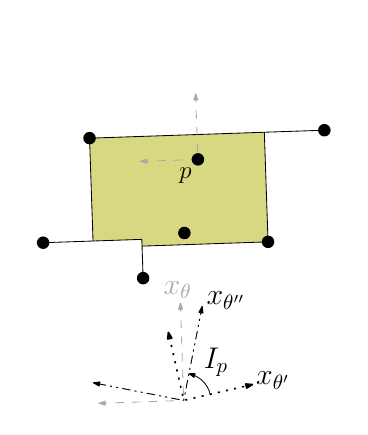}}
\end{center}
\caption{(a)~$I_p=[\theta',\theta'')$, together with $P$-free open $\theta'$- and $\theta''$-quadrants (rotations of the first quadrant) with apex~$p$. (b)~Situation for a $\theta\in I_p$ and $P$-free open $\theta$-quadrant with apex~$p$. (c)~Situation for a $\theta\notin I_p$ and non-$P$-free open $\theta$-quadrant with apex~$p$.}\label{fig:prel:interval}
\end{figure}

Every point $p\in \Vt$ is the apex of a $P$-free open $\theta$-quadrant in $\Wt$. In Figure~\ref{fig:prel:interval_ini} we show a point~$p$ that is in~$\Vt$ for all $\theta$ in the interval $I_p=[\theta',\theta'')$.
The endpoints of $I_p$ mark the \emph{in} and \emph{out} events of $p$; i.e., the values of $\theta$ when $p$ starts and stops being in $\Vt$. Since $P$ is in general position, each of its points can have at most three intervals~$I_p$ for which it is a vertex of~$\rcht$.

The following Theorem~\ref{mantenimiento} is not stated explicitly as a result in~\cite{fitting}, but it appears as a main step of an algorithm for a fitting problem. The proof is obtained by first computing the set of intervals~$I_p$ for which a point $p\in P$ is maximal with respect to some $\theta$-quadrant (using a result from Avis et al.~\cite{theta-maxima_1999}), then computing the ordered set of in- and out-events for points~$p$ while $\theta$ increases from~$0$ to~$2\pi$ (performing a line sweep with four lines to obtain the maximal points of~$P$ for each of the four $\theta$-quadrants). The reader is referred to~\cite{fitting} for further details.

\begin{theorem}[D\'{i}az-Ba\~{n}ez et al.~\cite{fitting}]\label{mantenimiento}
Computing and maintaining the $\theta$-rectilinear convex hull $\rcht$ while $\theta$ increases from~$0$ to~$2\pi$ can be done in optimal $O(n\log n)$ time and $O(n)$ space.
\end{theorem}

\subsection{Finding the value of \boldmath{$\theta$} for which \boldmath{$\rcht$} has minimum area}
\label{subsec:unorientedRCH}

\begin{figure}[ht]
  \centering
  {\includegraphics[scale=1.2]{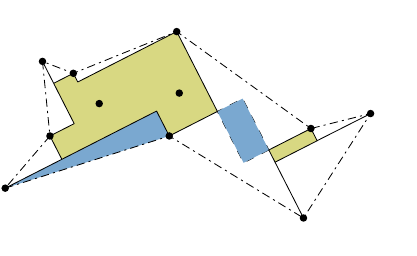}}
  \caption{\label{fig:prelim:area}Computing the area of $\rcht$. The polygonal region $\polygon$ is bounded by the dash-dotted line. A triangular and a rectangular overlap are shaded blue.}
\end{figure}

For a fixed value of $\theta$, we can compute the area of $\rcht$ using the fact that
\begin{equation}
  \label{eq:prelim:area}
  \area(\rcht) = \area(\polygon) - \area(\polygon \setminus \rcht),
\end{equation}
where $\polygon$ denotes the polygonal region having the points in $\Vt$ as vertices and an edge connecting two vertices if they are consecutive elements in $\Vt$; see \Cref{fig:prelim:area}. We will compute the area of $\polygon\setminus\rcht$ by decomposing it into two types of regions: (i)~the triangles defined by every pair of consecutive elements in $\Vt$, and (ii)~the rectangular overlaps between two triangles that make $\rcht$  disconnected; see again \Cref{fig:prelim:area}.

By Theorem~\ref{mantenimiento}, the triangles in~(i) above can be maintained in optimal $O(n\log n)$ time and $O(n)$ space. While $\theta$ increases from~$0$ to~$2\pi$, the set $\Vt$ of points on~$\partial(\rcht)$ changes at the values of~$\theta$ where a point of $P$ becomes (resp.\ is no longer) a vertex of $\rcht$. We call these angles \emph{in-events} (resp.\ \emph{out-events}).

In the rest of this subsection we will deal with the rectangles in~(ii), showing how to maintain the set $\St$ of rectangular overlaps that changes at \emph{overlap events} (resp.\ \emph{release events}), which are where such a rectangular overlap appears or disappears; see Figure~\ref{fig:prel:overlap}. It is important to note that there exist point configurations for which overlap and release events do not coincide with vertex events~\cite{bae_2009}, and hence the computations of $\Vt$ and~$\St$ are independent.

\subsubsection{Overlap and release events}\label{sec:urch:oet}

\begin{figure}[ht]
\begin{center}
  \subcaptionbox{\label{fig:prel:start_event}}
  {\includegraphics[width=0.29\textwidth]{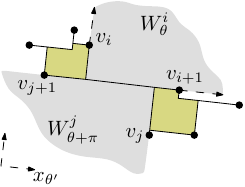}}
  \qquad
  \subcaptionbox{\label{fig:prel:middle}}
  {\includegraphics[width=0.29\textwidth]{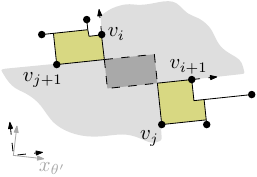}}
  \qquad
  \subcaptionbox{\label{fig:prel:stop_event}}
  {\includegraphics[width=0.29\textwidth]{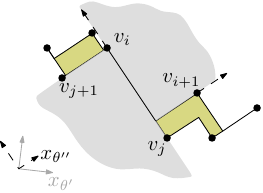}}
\end{center}
\caption{(a)~Overlap event (an overlap begins at angle~$\theta'$). (b)~Overlap that disconnects $\rcht$. (c)~Release event (an overlap ends at angle~$\theta''$).}\label{fig:prel:overlap}
\end{figure}

Let us label the points of $P$ in $V_\theta(P)$ as $v_1,\dots,v_m$, the order in which they appear as we traverse the boundary of $\rcht$ clockwise (recall Equation~\ref{eq:Vquadrants}). Let $W_\theta^i$ denote the $P$-free $\theta$-quadrant supported by two points $v_i,v_{i+1}\in V_{\theta}(P)$ (see Figure~\ref{fig:prel:start_event}), and proceed analogously for $V_{\theta+\frac{\pi}{2}}(P)$, $V_{\theta+\pi}(P)$, and $V_{\theta+\frac{3\pi}{4}}(P)$. The $P$-free $\theta$-quadrants obtained, which define the boundary of~$\rcht$, will be called \emph{extremal}.

$\theta$-quadrants and $(\theta+ \pi)$-quadrants will be called \emph{opposite quadrants}; see Figure~\ref{fig:prel:start_event}. When two opposite extremal quadrants intersect, as in Figure~\ref{fig:prel:middle} where $W_\theta^i \cap W_{\theta+\pi}^j\neq\emptyset$, we say that they \emph{overlap}, and we denote their intersection by~$S_{\theta}(i,j)$. For an angle $\theta$, the set $\St$ contains all intersections of opposite $\theta$- and $(\theta + \pi)$-quadrants.

Recall from the beginning of Subsection~\ref{subsection:rcht} that for a fixed value of~$\theta$, the set $\Vt$ of vertices can be computed in optimal $O(n\log n)$ time and $O(n)$ space. The set $\St$ of rectangular overlaps can be computed from $\Vt$ in $O(n)$ time. In the following, we show how to efficiently maintain the set $\St$ while $\theta$ increases from~$0$ to~$2\pi$.
We now justify why overlaps were defined above only for opposite extremal $\theta$-quadrants.

\begin{lemma}\label{lemma:overlap_oriented}
If two extremal $\theta$-quadrants have nonempty intersection, then they have to be opposite. When this happens, $\rcht$ becomes disconnected.
\end{lemma}

\begin{proof}
In any pair of non-opposite $\theta$-quadrants, one of them contains a ray parallel to a bounding ray of the other one. Since every extremal $\theta$-quadrant is supported by at least two points of~$P$ (recall that it defines part of the boundary of $\rcht$), then if a pair of non-opposite extremal $\theta$-quadrants has a nonempty intersection, one of them is not $P$-free, a contradiction. See again Figure~\ref{fig:prel:middle}.
\end{proof}

This property will be useful in the next two subsections, where we will show that the number of overlap and release events is linear, and we will obtain an optimal algorithm to compute them.

\subsubsection{The chain of arcs}\label{subsubsec:chain_arcs}

\begin{figure}[ht]
  \begin{center}
    \includegraphics[width=0.45\textwidth]{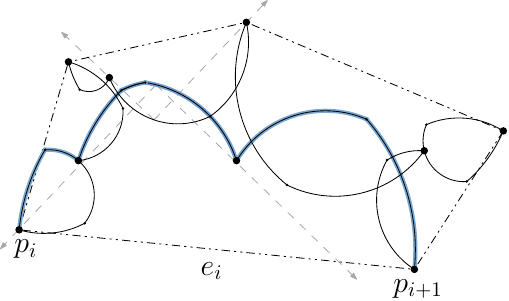}
    \quad
    \includegraphics[width=0.45\textwidth]{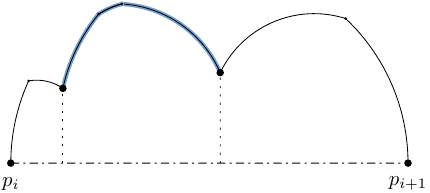}
  \end{center}
  \caption{Left: The arc-chain of $P$, highlighting the sub-chain associated with $e_i$. Right: Highlighted, a link of that sub-chain.}
  \label{fig:cor:arc-chain}
\end{figure}

Let the \emph{chain of arcs} of $P$, denoted by $\Ac$, be the curve composed of the points $a$ in the plane that are apexes of a $P$-free extremal $\theta$-quadrant $W^a$ for some $\theta\in [0,2\pi)$. Note that 
$W^a$ is supported by at least two points of $P$. The \emph{sub-chain} associated with an edge $e_i$ of $\partial(\ch)$ will be defined as the curve $A_{e_i}$ composed of those points $a$ such that $W^a$ intersects $e_i$. See \Cref{fig:cor:arc-chain}, left. This sub-chain $A_{e_i}$ is monotone with respect to $e_i$, since it is composed of arcs of circles, which have to be monotone in order for $W^a$ to intersect $e_i$, and two consecutive monotone arcs whose extremal $\theta$-quadrants intersect $e_i$ can only form a monotone curve. Finally, since a sub-chain may have vertices not belonging to $P$, we define a \emph{link} to be the part of a sub-chain that lies between two points of $P$. See \Cref{fig:cor:arc-chain}, right.

Note that if a pair of opposite extremal $\theta$-quadrants generates an overlapping region, then their apexes lie on intersecting links and the rectangular overlap lies in the area bounded by the intersection of the two links. See again \Cref{fig:cor:arc-chain}, left. Hence, in order to prove that the set~$\St$ of overlapping regions can be maintained in linear time and space, we will prove that the number of overlap and release events is linear. This is done by proving that the number of intersections between the links is linear.

\subsubsection{There is a linear number of intersections between links}\label{subsection:linear}

We now prove that the number of intersections between the links of the arc chain $\Ac$ is in $O(n)$. We first need a series of auxiliary lemmas.

\begin{lemma}\label{lem:linear:angle}
For any three points $a,b,c$ appearing from left to right on a link $\ell$, the angle $\angle abc \in [\frac{\pi}{2},\pi)$.
\end{lemma}

\begin{figure}[ht]
  \begin{center}
    \includegraphics[scale=1.1]{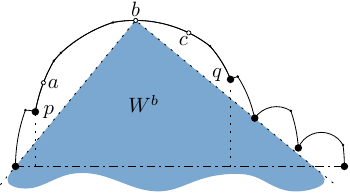}
    \qquad
    \includegraphics[scale=1.1]{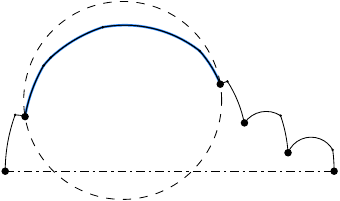}
  \end{center}
  \caption{Left: Illustration of \Cref{lem:linear:angle}. Right: A link disk.}
  \label{fig:linear:angulo_subcadena}
\end{figure}

\begin{proof}
Let $p$ and $q$ denote the endpoints of the link $\ell$. Since $p$ and $q$ are consecutive points of $P$ along the chain of arcs, then $b$ is not a point of $P$, $W^b$ being a $P$-free extremal $\theta$-quadrant. The fact that $\angle abc \geq\frac{\pi}{2}$
follows since $a,c$ are not in the interior of $W^b$, otherwise $W^b$ would not be $P$-free, either because some of $a,c$ is
actually an endpoint of $\ell$ (and thus a point of $P$) or because one of the points of $P$ supporting the extremal $\theta$-quadrants with apexes $a,c$ is in the interior of $W^b$. Suppose that $\ell$ is in the sub-chain $A_{e_i}$, for some edge $e_i$ of $\partial(\ch)$. The fact that $\angle abc< \pi$ follows from the orthogonal projections of $p$ and $q$ over $e_i$ being inside the intersection between $W^b$ and $e_i$. See \Cref{fig:linear:angulo_subcadena}, left.
\end{proof}

A crucial consequence of Lemma~\ref{lem:linear:angle} is the following. Let $a$ and $b$ denote two consecutive points along a
sub-chain $A_{e_i}$ for some edge $e_i$ of $\partial(\ch)$. The disk with the segment $\overline{ab}$ as diameter encloses the arcs of $A_{e_i}$ which are between $a$ and $b$. In particular, we are interested in the disks enclosing sub-chains and links. The former, that we call \emph{sub-chain disks}, have as diameter an edge of $\partial(\ch)$. The latter, that we call \emph{link disks}, have as diameter the segment connecting a pair of points of $P$ that are the endpoints of a link. See
Figure~\ref{fig:linear:angulo_subcadena}, right.

\begin{lemma}\label{lem:linear:link_inter}
Consider the link disks in two sub-chains $A_{e_i}$ and $A_{e_j}$. Suppose the link disk $D$ with smallest diameter is in $A_{e_i}$. Then $D$ can be intersected by at most five links in $A_{e_j}$.
\end{lemma}

\begin{proof}
Let $R$ be the strip bounded by the lines that orthogonally project $D$ over the edge $e_j$. Since $A_{e_j}$ is monotone with respect to $e_j$, only the part of $A_{e_j}$ in $R$ can intersect $D$; see \Cref{fig:linear:r}, left.

\begin{figure}[ht]
\begin{center}
\includegraphics[width=0.4\textwidth]{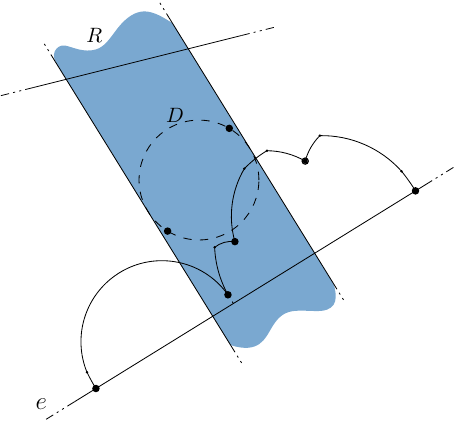}
\includegraphics[width=0.25\textwidth]{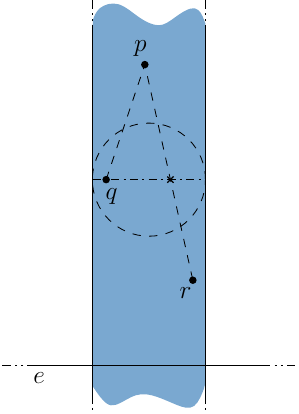}
\includegraphics[width=0.25\textwidth]{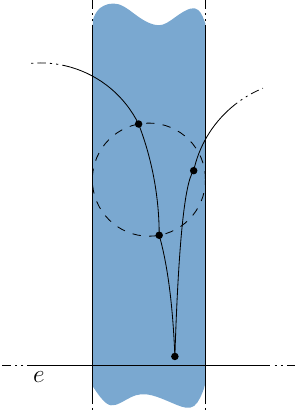}
\end{center}
\caption{Left: Only the part of the sub-chain inside $R$ can intersect $D$. Middle: There are no peaks at points of $P$ inside $R$. Right: At most $5$ links intersect $D$.}\label{fig:linear:r}
\end{figure}

If no arc of $A_{e_j}$ has endpoints inside $R$, then at most one link can intersect $D$. Otherwise, we claim that the sequence of points of $P$ in $A_{e_j}$ has no peaks inside $R$. Suppose for a contradiction there is a peak $p\in R$. Let $q$ and $r$ be the neighbors of $p$, $r$ being the one closer to $e_j$. Consider the line segment obtained by intersecting the strip $R$ and the line through $q$ parallel to $e_j$. This line segment is the diameter of a disk which does not contain the peak $p$, since the length of the segment $\overline{pq}$ equals the diameter of a link disk and hence, it has to be greater than the diameter of $D$, which equals the width of $R$. See \Cref{fig:linear:r}, middle. Therefore, the angle $\angle qpr$ is less than $\frac{\pi}{2}$, a contradiction with Lemma~\ref{lem:linear:angle}.

Since the sequence of points of $P$ in $A_{e_j}$ has no peaks inside $R$, it can have at most one valley inside $R$. Therefore, at most five links in $A_{e_j}$ can intersect $D$. See \Cref{fig:linear:r}, right.
\end{proof}

Let $n_i$ denote the number of points of $P$ in the sub-chain $A_{e_i}$.

\begin{lemma}\label{lem:linear:link_inter_linear}
Consider the set of links in two sub-chains $A_{e_i}$ and $A_{e_j}$. The number of pairs of intersecting links is in $O(n_i+n_j)$.
\end{lemma}

\begin{proof}
Let $\mathcal{L}=\{\ell_1,\ldots,\ell_m\}$ be the list of links in both $A_{e_i}$ and $A_{e_j}$, sorted in increasing order by the sizes of the diameters of their link disks. Let $\mathcal{L}_s= \mathcal{L}\setminus\{\ell_1,\ldots,\ell_s\}$, $s\geq 1$. It is easy to see by \Cref{lem:linear:link_inter} that $\ell_s$ is intersected by at most five links of $\mathcal{L}_s$, $s=1,\ldots,m-1$. Our result follows.
\end{proof}

\begin{lemma}\label{lemma:linear:linear}
Consider the set of links in two sub-chains $A_{e_i}$ and $A_{e_j}$. The number of intersection points between the links is in
$O(n_i+n_j)$.
\end{lemma}

\begin{figure}[ht]
  \centering
  \subcaptionbox{\label{fig:linear:link-intersection-2:1}}
  {\includegraphics[scale=1.1]{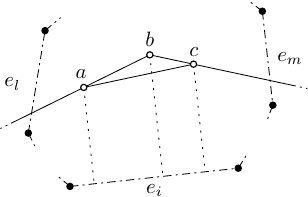}}
  \qquad{}
  \subcaptionbox{\label{fig:linear:link-intersection-2:2}}
  {\includegraphics[scale=1.1]{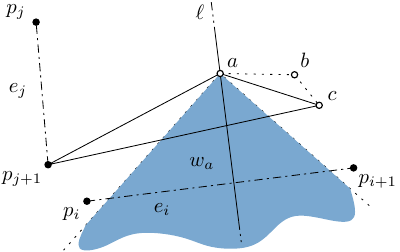}}
  \caption{\subref{fig:linear:link-intersection-2:1} Valid relative
    positions of the edge $e_j$,
    \subref{fig:linear:link-intersection-2:2} $c$ is not contained in
    $A_{e_j}$, since $\angle acp_{j+1}<\frac{\pi}{2}$.}
\label{fig:linear:link-intersection-2}
\end{figure}

\begin{proof}
Because of the monotonicity, we know that two links in the same sub-chain can intersect only at one of their endpoints. By
\Cref{lem:linear:link_inter_linear}, all we have to prove is that links from two different sub-chains intersect at most twice.

Suppose there are at least three intersection points $a,b,c$ between a link in $A_{e_i}$ and a link in $A_{e_j}$. Without loss of generality, assume that $a,b,c$ appear from left to right on the link in $A_{e_i}$. Note that they also appear from left to right on the link in $A_{e_j}$, since otherwise one of these points cannot belong to this link, as the three of them would form an angle either smaller than $\frac{\pi}{2}$ (\Cref{fig:linear:link-intersection:1}) or greater than $\pi$ (\Cref{fig:linear:link-intersection:2}), contradicting \Cref{lem:linear:angle}.

\begin{figure}[ht]
\centering
\subcaptionbox{\label{fig:linear:link-intersection:1}}
{\includegraphics[scale=1.1]{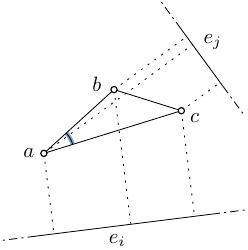}}
\qquad{}
\subcaptionbox{\label{fig:linear:link-intersection:2}}
{\includegraphics[scale=1.1]{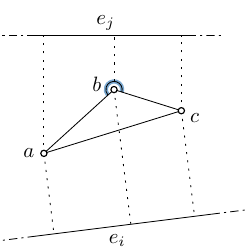}}
\caption{
      \subref{fig:linear:link-intersection:1} 
      $\angle cab <\frac{\pi}{2}$,
      \subref{fig:linear:link-intersection:2} 
      $\angle cba > \pi$.  }\label{fig:linear:link-intersection}
\end{figure}

Let $e_{l}$ and $e_{m}$ be respectively the edges of $\partial(\ch)$ intersected by the rays from $b$ passing through $a$ and $c$. While traversing the edges of $\partial(\ch)$ in the counterclockwise direction, $e_j$ lies either between $e_l$ and $e_i$, or between $e_i$ and $e_m$ (see \Cref{fig:linear:link-intersection-2:1}). Suppose that $e_j$ is between $e_l$ and $e_i$ (the second case follows in a similar way), and let $\ell$ be the line perpendicular to $e_{i}$ passing through~$a$. Since $W^a$ is a maximal wedge bounded by rays intersecting $e_{i}$, as in the proof of Lemma~\ref{lem:linear:angle}, $W^a$ does not contain any other point from the link in $A_{e_i}$ (see \Cref{fig:linear:link-intersection-2:2}). Note that $c$ and $p_{j+1}$ are on opposite sides of $\ell$ and are not contained in $W^a$, and thus $\angle p_{j+1}ac \ge\frac{\pi}{2}$ and $\angle acp_{j+1}<\frac{\pi}{2}$. Since $a,c,p_{j+1}$ appear from left to right on the link associated with $e_j$, we get from \Cref{lem:linear:angle} that $c$ cannot belong to the sub-chain~$A_{e_j}$.
\end{proof}

We are now ready to prove the main result of this section.

\begin{theorem}\label{thm:linear:linear}
The number of intersection points between all the links in the arc chain $\Ac$ is in~$O(n)$.
\end{theorem}

\begin{proof}
Consider the weighted graph constructed as follows. Each vertex represents a sub-chain, and an edge connects two vertices if their sub-chains intersect to each other. The weight of each edge is the number of intersection points between the sub-chains. Then, the total number of intersections equals the sum of the weights of all the edges of the graph. We prove next that this sum is in $O(n)$. Each point $p\in P$ can be in at most four sub-chain disks, since $p$ can be the apex of at most four $P$-free wedges of size $\frac{\pi}{2}$ (actually, of at most three if we consider general position). Thus, each point $p\in P$ can be in the intersection of at most ${{4}\choose{2}}=6$ pairs of sub-chain disks, therefore contributing to the weight of at most $6$ edges of the weighted graph. By Lemma~\ref{lemma:linear:linear}, the weight of each of these edges is linear on the number of points from $P$ contained in the corresponding sub-chains. Therefore, the sum of the weights of all the edges of the graph is linear on the total number of points in $P$.
\end{proof}

\subsubsection{Computing the sequence of overlap and release events}\label{subsection:table}

Next, we outline the algorithm to compute the sequence of overlap and release events.

\hspace{0.2cm}

\textsc{Event-sequence algorithm}

\begin{enumerate}

\item\label{enum:table:arc-chain} Compute the chain of arcs of $P$.

Each arc is stored as the points supporting the corresponding extremal $\theta$-quadrant and the angular interval defined by these points, called the \emph{tracing interval}. The elements in $\Ac$ are grouped by links.

  \begin{enumerate}
  \item At each insertion event, at most two arcs are generated and at most one arc is interrupted. Pointers are inserted from the interrupted arc to the arcs just generated. If an end of a new arc is a point in $P$, a new link is initialized with the respective arc.

  \item At each deletion event, at most one arc is generated, and at most two arcs are interrupted. One of the interrupted arcs will be always ending at a point in $P$. As before, pointers are set up from the interrupted arc to the newly created arcs.
  \end{enumerate}

\item\label{enum:table:colored-arc-chain} Color the arcs.

  Traverse $\Ac$ in such way that the vertices of $\partial(\ch)$ are visited in counterclockwise order, while assigning the following colors to each arc: red if its subchain corresponds to an edge in the upper chain $\partial(\ch)$ and blue otherwise (see \Cref{fig:table:arc-chain}). Note that regardless of the value of $\theta$, a pair of extremal $\theta$-quadrants intersecting an edge in the upper chain (resp.\ lower chain) of $\partial(\ch)$ are not opposite to each other. Then if there is an intersection between links of the same color, such an intersection does not correspond to an overlap between extremal $\theta$-quadrants.

\begin{figure}[ht]
\centering
\includegraphics[scale=1]{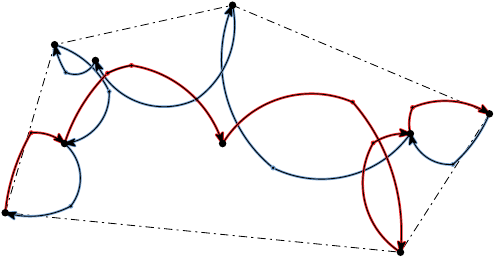}
\caption{The colored chain of arcs of $P$.}
\label{fig:table:arc-chain}
\end{figure}

\item\label{enum:table:arc-intersections} Identify bichromatic intersecting links.

  Note that the largest possible arc is a semicircle, and therefore, any arc can be partitioned into at most three segments to get a set of curves monotone with respect to an arbitrary direction. The arcs in $\Ac$ can thus be transformed into a set $\mathcal{A}^{\prime}(P)$ of curves monotone with respect to the same direction. The Bentley and Ottmann plane sweep algorithm~\cite{bentley_1979a} can then be applied to $\mathcal{A}^{\prime}(P)$ to compute the intersection points between arcs. From among these points, we distinguish those belonging to  pairs of arcs of different color. Pointers to the links containing the arcs involved in each intersection are set up, so that we can obtain the set of all pairs of intersecting links of different color in $\Ac$.

\item\label{enum:table:intersecting-links} Compute the sequence of overlap and release events.

Given two points $u$ and $v$, let $C(u,v)$ be the semicircle whose diameter is the segment joining $u$ to $v$, such that it starts in $u$, moves clockwise, and ends in $v$.
Consider two extremal $\theta$-quadrants, denoted as $\qd$ and $\qd[r][s]$, and a pair of arcs $\arc\in C(p,q)$ and $\arc[c][d]\in C(r,s)$ together with their corresponding tracing intervals $\left(\alpha_a,\alpha_b\right)$ and $\left(\alpha_c,\alpha_d\right)$. See \Cref{fig:urch:oet:ov}. We say that $\arc$ and $\arc[c][d]$ \emph{admit overlapping $\theta$-quadrants}, if $\qd[][][\varphi]$ and $\qd[r][s][\psi]$ overlap for some $\varphi\in\left(\alpha_a,\alpha_b\right)$ and $\psi\in\left(\alpha_c,\alpha_d\right)$.

\begin{figure}[ht]
\centering
\subcaptionbox{\label{fig:urch:oet:ov:1}}
{\includegraphics[width=0.45\textwidth]{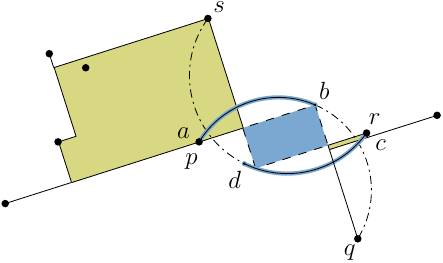}}
\qquad
\subcaptionbox{\label{fig:urch:oet:ov:2}}
{\includegraphics[width=0.45\textwidth]{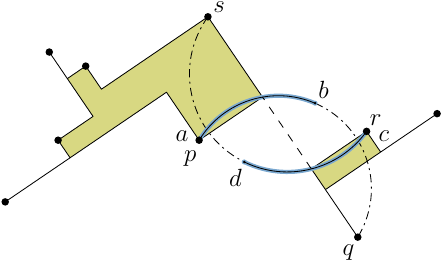}}
\caption{(a)~The arcs $\arc$ and $\arc[c][d]$ (highlighted) admit
  overlapping $\theta$-quadrants. (b)~Release event of the
  corresponding overlapping region.}\label{fig:urch:oet:ov}
\end{figure}

  Assume that $\arc$ and $\arc[c][d]$ admit overlapping of opposite $\theta$-quadrants and without loss of generality suppose that $p$ precedes $q$ in $\Vt$ for all $\theta \in (\alpha_a,\alpha_b)$, and that $r$ precedes $s$ for all $\theta \in (\alpha_c,\alpha_d)$. It is not hard to see that since the extremal $\theta$-quadrants $\qd$ and $\qd[r][s]$ are opposite, $\left(\alpha_a,\alpha_b\right) \cap \left(\alpha_c+\pi,\alpha_d+\pi\right)$ is not empty and during this interval, the ray of $\qd$ passing through $p$ (resp.\ $q$) is parallel to the ray of $\qd[r][s]$ passing through $r$ (resp.~$s$). Note that $q$ and $s$ lie on different sides of the line $\ell_{p,r}$ passing through $p$ and $r$, for otherwise $\qd\cap \qd[r][s]$ could not be $P$-free. For the same reason, the points $p,r$ lie on opposite sides of $\ell_{q,s}$, and therefore the line segments $\overline{pr}$ and $\overline{qs}$ intersect. It is easy to see that this intersection is contained in the overlapping region generated by $\qd$ and $\qd[r][s]$, and thus we have that $\overline{pr}\cap\overline{qs}\subset \qd\cap \qd[r][s]$. Note that the angular interval of maximum size where $\qd$ and $\qd[r][s]$ may overlap. We call this a \emph{maximum overlapping interval}, and it is bounded by the orientations where $\X$ is parallel to $\overline{pr}$ and where $\Y$ is parallel to $\overline{qs}$.

\begin{observation}\label{lem:cor:conditions}
The arcs $\arc$ and $\arc[c][d]$ admit overlapping $\theta$-quadrants if and only if $\qd$ and $\qd[r][s]$ define a maximum overlapping interval $(\theta_1,\theta_2)$, and
$$(\theta_1,\theta_2)\cap\left(\alpha_a,\alpha_b\right)\cap\left(\alpha_c+\pi,
\alpha_d+\pi\right)\neq\emptyset.$$
\end{observation}

Let $\left\langle\arc[a_1][a_2],\arc[a_2][a_3],\dots,\arc[a_k][a_{k+1}]\right\rangle$ be the set of arcs for all $\theta\in [0,2\pi)$, where $k=O(n)$, labeled while traversing $\Ac$ in such way that the vertices of $\partial(\ch)$ are visited in counterclockwise order. We denote by $\larc$ the subsequence $\left\langle (a_u,a_{u+1}),\dots,(a_v,a_{v+1}) \right\rangle$ of consecutive arcs in $\Ac$ forming a link. Note that the extremal intervals of the arcs in $\larc$ define a sequence $\left\langle \alpha_{a_u},\dots,\alpha_{a_{v+1}}\right\rangle$ of
increasing angles. See Figure~\ref{fig:link-consecutive}.

\begin{figure}[ht]
\centering
\subcaptionbox{\label{fig:link-consecutive:1}}
{\includegraphics{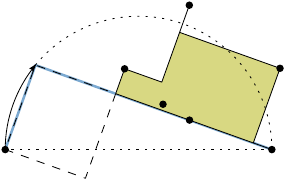}}
\qquad
\subcaptionbox{\label{fig:link-consecutive::2}}
{\includegraphics{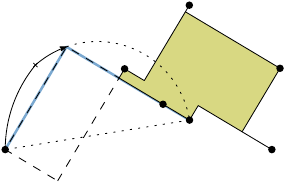}}
\qquad
\subcaptionbox{\label{fig:link-consecutive:3}}
{\includegraphics{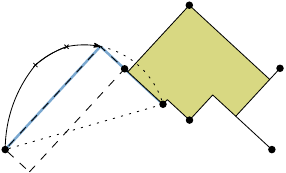}}
\qquad
\subcaptionbox{\label{fig:link-consecutive:4}}
{\includegraphics{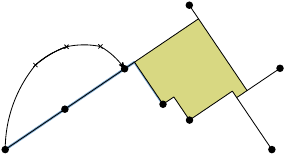}}
\caption{The extremal intervals in the same link define a sequence of
  increasing angles.}
\label{fig:link-consecutive}
\end{figure}

By Observation~\ref{lem:cor:conditions}, it is easy to see that we can compute the overlapping regions generated by the arcs belonging to a pair $\larc$ and $\larc[s][t]$ of intersecting links, using a procedure similar to that of merging two sorted lists where the lists are $\left\langle \alpha_{a_u},\dots,\alpha_{a_{v+1}}\right\rangle$ and
$\left\langle \alpha_{a_s}+\pi,\dots,\alpha_{a_{t+1}}+\pi\right\rangle$ and their corresponding arc sequences. The intersection between a pair of non-consecutive maximal intervals in the merged list is empty. These pairs can be ignored, as they do not meet the conditions of Observation~\ref{lem:cor:conditions}, and therefore at most a linear number of pairs of arcs in $\larc$ and $\larc[s][t]$ admit overlapping $\theta$-quadrants.

Let $\link[u][v]$ and $\link[s][t]$ be two intersecting links containing $n_{u,v}=u-v+1$ and $n_{s,t}=t-s+1$ arcs respectively. At most $O(n_{u,v}+n_{s,t})$ pairs of arcs admit overlapping $\theta$-quadrants. The overlapping regions generated by the admitted extremal $\theta$-quadrants can be computed using $O(n_{u,v}+n_{s,t})$ time and space.
\end{enumerate}

By \Cref{mantenimiento}, Step~\ref{enum:table:arc-chain} takes $O(n\log n)$ time and $O(n)$ space, since a constant number of additional operations are performed at each event while traversing the vertex event sequence.
Step~\ref{enum:table:colored-arc-chain} takes $O(n)$ time and space, since the number of arcs in $\Ac$ is linear in the number of elements in $P$. The time required to compute $\mathcal{A}(P)$ is in $O(n)$. By \Cref{thm:linear:linear}, the Bentley and Ottmann~\cite{bentley_1979a} plane sweep processes on $\mathcal{A}(P)$ take $O(n\log n)$ time and $O(n)$ space. Additional linear time is needed to discriminate, from among the resulting intersection points, those belonging to intersections of a red and a blue link,  so Step~\ref{enum:table:arc-intersections} requires a total of $O(n\log n)$ time and $O(n)$ space. Finally, from Bae et al.~\cite[Lemma 5]{bae_2009} we know that there are $O(n)$ overlap and release events, and thus, Step~\ref{enum:table:intersecting-links} can be done in $O(n\log n)$ time and $O(n)$ space. Therefore, we have the following result.

\begin{theorem}\label{thm:table:overlap}
The sequence of $O(n)$ overlap and release events of~$\rcht$ while~$\theta$ increases from~$0$ to~$2\pi$ can be computed in $O(n\log n)$ time and $O(n)$ space.
\end{theorem}

\subsubsection{Sweeping the sequence of overlap and release events}\label{subsection:maintain}

We now store $\St$ in a hash table, using tuples with the points supporting the overlapping $\theta$-quadrants as keys, in the same order as they are encountered while traversing $\mathcal{V}_\theta(P)$. For an example, the overlapping region in Figure~\ref{fig:urch:oet:ov:2} would be stored in $\St$ using the tuple~$(p,q,r,s)$ as key.

\begin{figure}[ht]
\begin{center}
\includegraphics[width=0.6\textwidth]{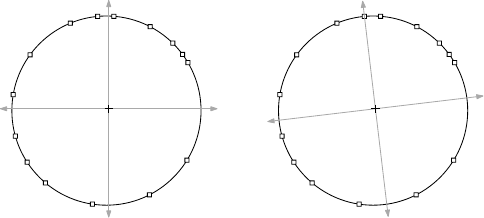}
\end{center}
\caption{\label{fig:overlap_events_sweep} Left: Representation of overlap and release events and $0$-quadrants.
Right: Simultaneous rotation of the four $\theta$-quadrants, stopping when one of their defining rays passes over an overlap event.}
\end{figure}

We also store the sequence of overlap and release events as points on a unit circle $[0,2\pi)$, over which we represent the $\theta$-quadrants. See Figure~\ref{fig:overlap_events_sweep}, left. Then we  simultaneously rotate the four $\theta$-quadrants counterclockwise around the center, stopping when one of their defining rays passes over a point representing an overlap event, and  update $\St$ accordingly; see Figure~\ref{fig:overlap_events_sweep}, right. It is easy to see that at any fixed value of $\theta$, there are $O(n)$ overlapping regions in~$\rcht$, which can be computed in linear time from $\mathcal{V}_\theta(P)$.

\begin{theorem}\label{thm:maintain:maintain}
Using the sequence of overlap and release events, the set $\St$ can be maintained in $O(n)$ time and $O(n)$ space as $\theta$ increases from $0$ to $2\pi$.
\end{theorem}

\subsection{Minimum area}\label{sec:area}

In this section we adapt the results from Bae et al.~\cite{bae_2009} to compute the value of $\theta$ that minimizes the area of $\rcht$ in optimal $O(n\log n)$ time and $O(n)$ space.

Let $(\alpha,\beta)$ be an angular interval in $[0,2\pi)$ containing no events. Extending~\Cref{eq:prelim:area}, we express the area of $\rcht$ for any $\theta \in (\alpha, \beta)$ as
\begin{equation}
  \label{eqn:area:area}
  \area(\rcht)
  \ =\ \area(\polygon)
  \ -\ \sum_{j} \area(\triangles[][j])
  \ +\ \sum_{k} \area(\squares[][k]).
\end{equation}

The term $\triangles[][j]$ denotes the triangular region bounded by the line through two consecutive vertices $v_j,v_{j+1}\in \Vt$, the line through $v_{j}$ parallel to the rotation of the $x$-axis, and the line through $v_{j+1}$ parallel to the rotation of the $y$-axis.~Finally, the term $\squares[][k]$ denotes the $k$-th overlapping region in $\St$.

We now show that for any particular value of $\theta$, \Cref{eqn:area:area} can be computed in linear time and, as $\theta$ increases from $0$ to $2\pi$, a constant number of terms need to be updated at each event, regardless of its type.

\subsubsection{The polygonal region}

\begin{figure}[ht]
\centering
\subcaptionbox{\label{fig:area:polygon:1}}
{\includegraphics{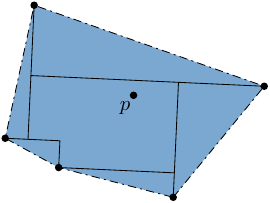}}
\hspace{1.5cm}
\subcaptionbox{\label{fig:area:polygon:2}}
{\includegraphics{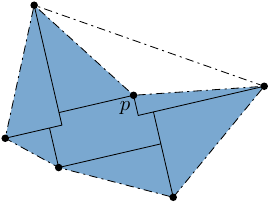}}
\caption{\subref{fig:area:polygon:1} The point $p$ is about to become a vertex. \subref{fig:area:polygon:2} After the insertion event of $p$, the area of the white triangle needs to be subtracted from $\area(\polygon)$.}
\label{fig:area:polygon}
\end{figure}

At any fixed value of $\theta$, the area of $\polygon$ can be computed from $\Vt$ in $O(n)$ time.~The term $\area(\polygon)$ changes only at vertex events.~These changes can be handled in constant time: at an \emph{in-event} (resp.\ \emph{out-event}), the area of a triangle needs to be subtracted (resp.\ added) to the previous value of $\area(\polygon)$.~See \Cref{fig:area:polygon}.

\subsubsection{The triangular regions}

\begin{figure}[ht]
\centering
\subcaptionbox{\label{fig:area:triangle:1}}
{\includegraphics[scale=1.2]{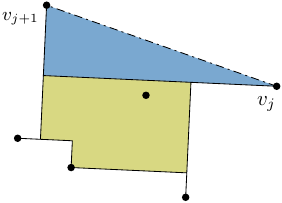}}
\hspace{1cm}
\subcaptionbox{\label{fig:area:triangle:2}}
{\includegraphics[scale=1.2]{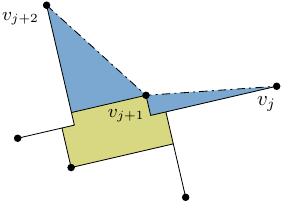}}
\caption{Updating the term $\sum_j\area(\triangles[][j])$. At an insertion event, \subref{fig:area:triangle:1} at most one triangle has to be removed, and \subref{fig:area:triangle:2} at most two triangles have to be added.}\label{fig:area:triangle}
\end{figure}

According to Bae et al.~\cite{bae_2009}, the area of $\triangles[][j]$ can be expressed as
\begin{equation}\label{eqn:area:triangles:1}
  \area(\triangles[][j]) = b^2_j \cdot \cos (c_j + (\theta - \alpha)) \cdot \sin (c_j + (\theta - \alpha)),
\end{equation}
where $b^2_j$ and $c_j$ are constant values that depend on the coordinates of the vertices supporting the $\theta$-quadrant that bounds $\triangles[][j]$.~\Cref{eqn:area:triangles:1} reduces to
\begin{align}
  \label{eqn:area:triangles:2}
  \area(\triangles[][j]) &= \frac{1}{2} b^2_j \cdot \sin 2(c_j + (\theta - \alpha)) \nonumber \\
                         &= \frac{1}{2} b^2_j \cdot \left[ \, \sin (2c_j) \cdot \cos 2(\theta - \alpha) +  \cos (2c_j) \cdot \sin 2(\theta - \alpha) \, \right] \nonumber \\
                         &= B_j \cdot \cos 2(\theta - \alpha) + C_j \cdot \sin 2(\theta - \alpha),
\end{align}
where $B_j = \frac{1}{2} b^2_j \cdot \sin (2c_j)$ and $C_j=\frac{1}{2} b^2_j \cdot \cos (2c_j)$. Equation~\ref{eqn:area:triangles:2} can be computed in constant time. There are $O(n)$ triangles, since the number of vertices in~$\Vt$ is linear. Thus, for any value of~$\theta$, the term $\sum_j\area(\triangles[][j])$ can be computed in $O(n)$ time. In an insertion event, the term for one triangle is removed from $\sum_j\area(\triangles[][j])$, and as a vertex supports at most two extremal $\theta$-quadrants, the terms of at most two triangles are added. The converse occurs for deletion events. The term
$\sum_j\area(\triangles[][j])$ is not affected by overlap or release events. See \Cref{fig:area:triangle}.

\subsubsection{The overlapping regions}

According to Bae et al.~\cite{bae_2009} the area of the $k$-th overlapping region can be expressed as
\begin{align}\label{eqn:area:ov:1}
  \area(\squares[][k]) &= \left| (x_i - x_j) \cos (\theta - \alpha) + (y_i - y_j) \sin (\theta - \alpha) \right|\\ \nonumber
                       & \times
                         \left| (y_{i+1} - y_{j+1}) \cos (\theta - \alpha) + (y_{i+1} - y_{j+i}) \sin (\theta - \alpha) \right|,
\end{align}
where $(x_i,y_i)$ are the coordinates of the vertex $p_i$, and similarly for $p_i,p_{i+1},p_j,$ and $p_{j+1}$. These four points are the vertices that support the overlapping $\theta$-quadrants that generate $\squares[][k]$. Simplifying  Equation~\ref{eqn:area:ov:1}, the area of the $k$-th overlapping region can be reduced to
\begin{equation}\label{eqn:area:ov}
  \area(\squares[][k])=B_k + C_k \cos 2(\theta - \alpha) + D_k \sin 2(\theta - \alpha),
\end{equation}
where $B_k$, $C_k$, and $D_k$ are constants that depend on the coordinates of the vertices supporting the overlapping
$\theta$-quadrants that generate $\squares[][k]$.~\Cref{eqn:area:ov} can be computed in constant time, and since there are $O(n)$ overlapping regions in $\St$, for any fixed $\theta$,  $\sum_{k}\area(\squares[][k])$ can be computed in $O(n)$ time. Overlap or release events require the term of a single overlapping region to be added to or deleted from $\sum_{k}\area(\squares[][k])$.~As a vertex supports at most two extremal $\theta$-quadrants, at an overlap event the terms of at most two overlapping regions are added or deleted.

Before describing the minimum area algorithm, we need the next three important properties of $\area(\rcht)$. First, from Lemma 4 in Bae et al.~\cite{bae_2009}, there are configurations of points such that the angles corresponding to optimal solutions are not necessarily the angles of vertex in-events or out-events; i.e., the value of $\theta$ for which $\area(\rcht)$ is minimum in $(\alpha,\beta)$ will be such that $\alpha<\theta<\beta$.~Second, \Cref{eqn:area:area} has $O(n)$ terms for any
$\theta\in (\alpha,\beta)$, and thus it can be reduced to
\begin{equation}
  \label{eqn:area:reduced-area}
  \area(\rcht)=C+D \cos 2(\theta -\alpha) +E \sin 2(\theta -\alpha)
\end{equation}
in $O(n)$ time.~The terms $C$, $D$ and $E$ denote constants resulting from summing together the constant values in $\area(\polygon)$ and in \Cref{eqn:area:triangles:2,eqn:area:ov}.~Finally, as
\Cref{eqn:area:reduced-area} has a constant number of inflection points in $\idos$, a constant number of operations suffice to obtain the value of $\theta$ that minimizes $\area(\oht)$ in $(\alpha,\beta)$. Note that at this point we can also ask for the value of $\theta$ that maximizes $\area(\oht)$ in $(\alpha,\beta)$, and that the maximum can in fact also take place at either of the extremes $\alpha$ or $\beta$.

\subsubsection{The search algorithm}\label{subsubsec:search_algorithm}

We now outline our algorithm to compute the angle~$\theta$ for which $\rcht$ has minimum area.

\begin{enumerate}
\item \label{area:step_1} \textbf{Compute the sequence of events.}

  Compute the sequence of vertex in-events and out-events as described in \Cref{subsection:rcht}, and the sequence of overlap and release events as described in \Cref{subsection:table}. Merge both sequences into a single circular sequence of angles $\left<\theta_1,\theta_2,\dots,\theta_{m-1},\theta_{m},\theta_1\right>$, $m\in O(n)$, which we can represent in a circular table as in Figure~\ref{fig:overlap_events_sweep}. Clearly, as $\theta$ increases from $0$ to $2\pi)$, the relevant features of $\rcht$  we are interested in (area, number of vertices, etc.) remain unchanged during each interval $(\theta_{i},\theta_{i+1})$, and each angle $\theta_i$ is an in-, out-, overlap-, or release-event.

\item \label{area:step_2} \textbf{Initialize the angular sweep.}

  Consider four $\theta$-quadrants, and suppose that the first (counterclockwise) defining ray of the first $\theta$-quadrant
  intersects the interval $(\theta_1,\theta_2)$. Compute the sets $\mathcal{V}_{\theta_1}$ and $\mathcal{S}_{\theta_1}$ as before, and calculate the $\area(\rcht)$ for $\theta\in[\theta_1,\theta_2)$ using \Cref{eqn:area:reduced-area}. Compute the constant values in this equation under the restriction $\theta\in [\theta_1,\theta_2)$. Optimize the resulting equation to compute the angle $\theta_{min}$ (resp.\ $\theta_{max}$) of minimum (resp.\ maximum) area.

\item \label{area:step_3} \textbf{Perform the angular sweep.}

  Simultaneously rotate the four $\theta$-quadrants as in \Cref{subsection:maintain}. During the sweeping process, update $\Vt$ and $\St$ as explained above. Additionally, at each event:
  \begin{enumerate}
  \item \label{area:step_3_1} Update \Cref{eqn:area:reduced-area} by adding or subtracting terms as explained above.
  \item \label{area:step_3_2} Optimize the updated \Cref{eqn:area:reduced-area} to obtain the local
  		 angle of minimum (resp.\ maximum) area, and replace $\theta_{min}$ (resp.\ $\theta_{max}$)
		 if a new minimum (resp.\ maximum) is obtained. 
  \end{enumerate}
\end{enumerate}

From Theorems~\ref{mantenimiento} and~\ref{thm:table:overlap}, computing the sequences of vertex, and overlap and release events takes $O(n\log n)$ time and $O(n)$ space. As both sequences have $O(n)$ events, we require linear time to merge them into the sequence of events, and thus Step~\ref{area:step_1} takes a total of $O(n\log n)$ time and $O(n)$ space. At Step~\ref{area:step_2}, $\Vt$ can be computed in $O(n\log n)$ time and $O(n)$ space (see~\cite{kung_1975}), and $\St$ can easily be computed from $\Vt$ in linear time. An additional linear time is required to obtain \Cref{eqn:area:reduced-area}, while $\theta_{min}$ (resp.\ $\theta_{max}$) can be computed in constant time. This gives a total of $O(n\log n)$ time and $O(n)$ space. Finally, by Theorems~\ref{mantenimiento} and~\ref{thm:maintain:maintain}, maintaining $\Vt$ and $\St$ requires $O(n\log n)$ time and linear space, respectively, for each. Step~\ref{area:step_3_1} and Step~\ref{area:step_3_2} are repeated $O(n)$ times (one per event in the sequence), and as we described above, each repetition takes constant time. Therefore Step~\ref{area:step_3} takes $O(n\log n)$ time and $O(n)$ space. Notice that after the sweeping process is finished, in a further $O(n\log n)$ time and $O(n)$ space we can compute both $\rcht$ and $\area(\rcht)$ for the angle $\theta_{min}$ (resp.\ $\theta_{max}$), giving the minimum (resp.\ maximum) area.

The algorithm is optimal since given $\rcht$, we compute $\mathcal{CH}(\rcht)=\mathcal{CH}(P)$ in linear time, and it is known that computing the convex hull of a set of $n$ points in the plane has an $\Omega(n\log n)$ time lower bound~\cite{preparata_1985}. Thus, we have the following.

\begin{theorem}\label{thm:unoriented_rch}
Computing $\rcht$ for the value of $\theta\in [0,2\pi)$ such that $\rcht$ has minimum (or maximum) area can be done in optimal $O(n\log n)$ time and $O(n)$ space.
\end{theorem}

\section{\boldmath{$\os$}-hull of a point set}\label{sec:3}

As mentioned in Section~\ref{sec:intro}, orthogonal convexity can be generalized to consider a finite set $\os$ of~$k$ different lines passing through the origin. A set is thus said to be \emph{$\os$-convex} if its intersection with any line parallel to an element of $\os$ is either connected or empty. Following the lines of Section~\ref{sec:rectilinear_hull}, we study now the $\os$-convex hull of a set of points.

\subsection{Definitions}

\begin{figure}[ht]
  \begin{center}
  \includegraphics{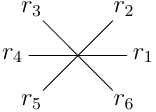}
  \\[1em]
  {\includegraphics{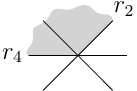}}
  \hfill
  {\includegraphics{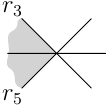}}
  \hfill
  {\includegraphics{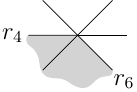}}
  \hfill
  {\includegraphics{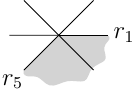}}
  \hfill
  {\includegraphics{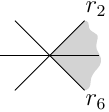}}
  \hfill
  {\includegraphics{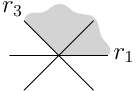}}
  \end{center}
  \caption{\label{fig:o-wedges}Top: A set $\os$ with $k=3$. Bottom: From left to right, the corresponding {$W_{{i+1},{i+k}}$} wedges for $i=1,\dots,2k$.}
\end{figure}

Let us assume that the lines in $\os$ are sorted and labelled according to their slopes as $\ell_1,\dots,\ell_k$. The origin splits each $\ell_i$ into two rays $r_i$ and $r_{i+k}$. Given two indexes $i$ and $j$, we define the wedge $W_{i,j}$ to be the open region spanned as we rotate~$r_i$ counterclockwise until it reaches $r_j$. A translation of a $W_{i,j}$ wedge will be called a $W_{i}^{j}$ wedge, and it is \emph{$P$-free} if it does not contain any point of $P$. Of particular interest to us is the set of $W_{i+1}^{i+k}$ wedges, $i=1,\ldots,2k$, which we will call the set of \emph{$\os$-wedges}. See Figure~\ref{fig:o-wedges}.

\begin{figure}[ht]
\begin{center}
\includegraphics[scale=0.8]{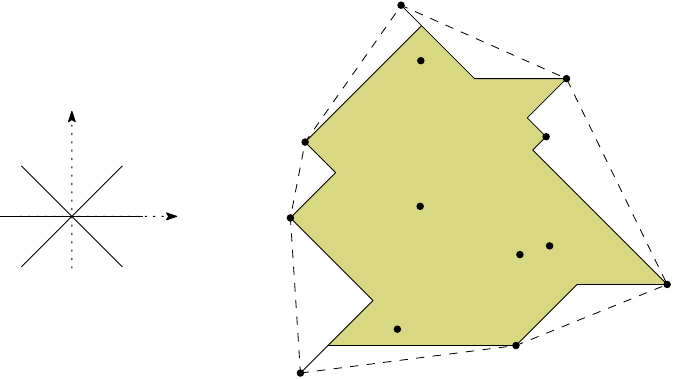}
\end{center}
\caption{\label{fig:o-hull} The set $\os$ in Figure~\ref{fig:o-wedges} and the $\os$-hull $\oh$ for a point set $P$.}
\end{figure}

Let $\mathcal W^i$ be the set of all the $P$-free $W_{i+1}^{i+k}$ wedges. Thus, by analogy with the orthogonal case in Section~\ref{sec:rectilinear_hull}, the \emph{$\os$-hull} of $P$ is
\[
  \oh =\mathbb{R}^{2} \setminus\bigcup_{i=1}^{2k}\W^i.
\]
See Figure~\ref{fig:o-hull} for an example.

\begin{figure}[ht]
\begin{center}
\includegraphics[width=0.24\textwidth]{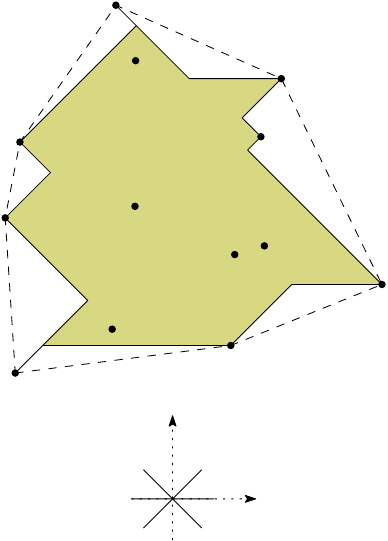}
\includegraphics[width=0.24\textwidth]{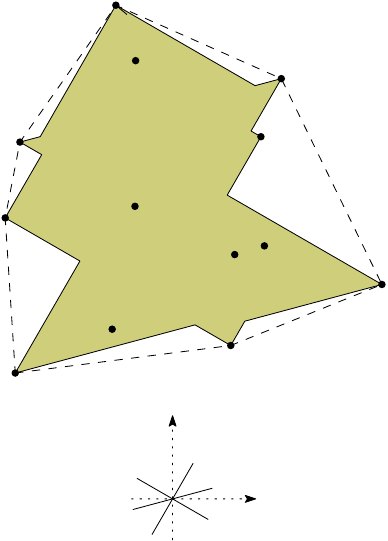}
\includegraphics[width=0.24\textwidth]{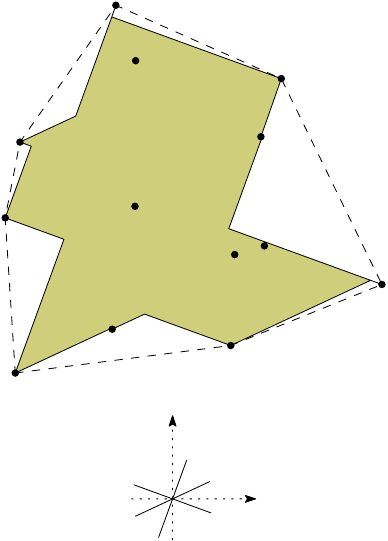}
\includegraphics[width=0.24\textwidth]{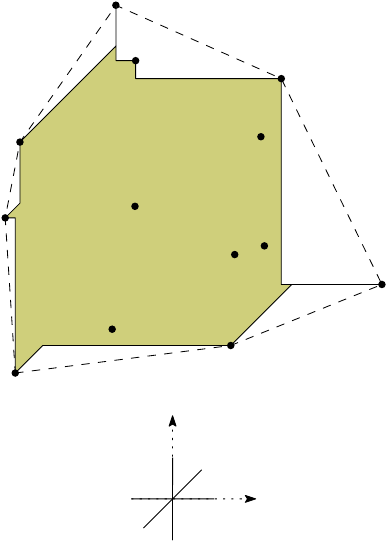}
\end{center}
\caption{\label{fig:o-hull-rotating} Changes on the $\ost$-hull $\oht$ as~$\theta$ changes.}
\end{figure}

Let $\ost$ be the set of lines obtained after rotating the elements of $\os$ by an angle $\theta$. Clearly, the $\ost$-hull of $P$, denoted as $\oht$, changes as $\theta$ goes from~$0$ to~$2\pi$. We will denote the resulting set as $\mathcal W^i_{\theta}$, so $\oht$ is now defined as
\[
  \oht =\mathbb{R}^{2} \setminus\bigcup_{i=1}^{2k}\W_\theta^i.
\]
 See Figure~\ref{fig:o-hull-rotating} for an example.

\subsection{Computing \boldmath{$\oh$}}\label{subsec:oriented_Ohull}

In this subsection we obtain an optimal $O(n\log n)$ time and $O(n)$ space algorithm to compute the $\os$-hull $\oh$ of a set~$P$ of $n$ points.

\subsubsection{Computing the vertices}\label{subsubsec:oriented_Ohull}

\begin{figure}[ht]
  \begin{center}
    \includegraphics[width=0.9\textwidth]{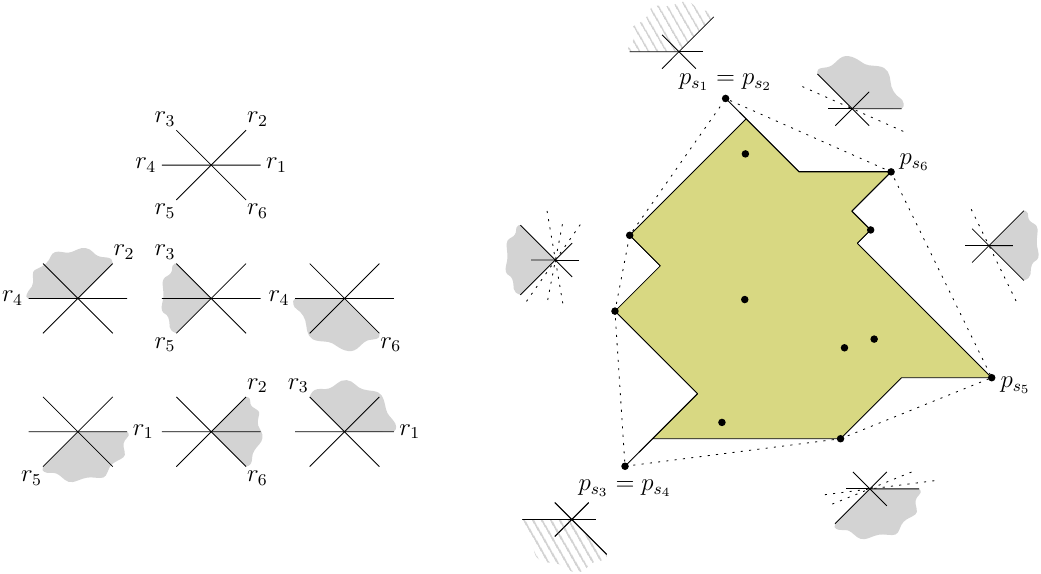}
  \end{center}
  \caption{\label{fig:stabbing}Left: Recalling Figure~\ref{fig:o-wedges}. Right: The $\os$-hull $\oh$ in Figure~\ref{fig:o-hull}, showing which edges of $\partial(\ch)$, if any, are intersected by wedges in each~$\W^i$. Note that the wedges in $\W^1$ and $\W^3$, shaded with hatched lines, do not intersect the interior of $\ch$. For $p_1$ as the uppermost point and labelling the vertices of~$\partial(\ch)$ counterclockwise, the stabbing intervals are $[1,4]$, $[4,6]$, $[6,7]$, and~$[7,1]$.}
\end{figure}

For each $r_i$, first compute the directed line parallel to $r_i$ which supports (the pre-computed) $\ch$ leaving $P$ on its right side. Suppose without loss of generality that each of these lines intersects $\ch$ at a single point, labelled $p_{s_i}$, $i=1,\ldots,2k$. Note that it is not necessarily true that $p_{s_i}$ is different from $p_{s_{i+1}}$; see Figure~\ref{fig:stabbing}. Thus, $p_{s_1},p_{s_2},\dots,p_{s_{2k}}$ are vertices of the boundary of the $\os$-hull, $\partial (\oh)$, labelled as we meet them in counterclockwise order. Note also that these~$p_{s_i}$ might not cover all the vertices of~$\partial (\oh)$; see Figure~\ref{fig:stabbing}.

Because of the definition $\oh=\mathbb{R}^{2}\setminus\cup_{i=1}^{2k}\W^i$, we need to compute $\partial(\W^i)$, and this requires knowing when a wedge in $\W^i$ can intersect the interior of $\ch$. It is easy to see that there are wedges in $\W^i$ that intersect the interior of $\ch$ if and only if $p_{s_i}\neq p_{s_{i+1}}$, and that any wedge in $\W^i$ intersecting the interior of $\ch$ necessarily does so by intersecting an edge of $\partial(\ch)$ whose endpoints $p_j,p_{j+1}$ are such that $s_i\le j,j+1<s_{i+1}$. See Figure~\ref{fig:stabbing}. Abusing the notation, let us denote by $[s_i,s_{i+1}]$ the closed interval of those indices of vertices on~$\partial(\ch)$ between~$s_i$ and~$s_{i+1}$, and let us call them the \emph{stabbing interval} of $\W^i$. See the caption of Figure~\ref{fig:stabbing}.

\begin{observation}\label{obs:convhull}
If $s$ belongs to the stabbing interval $[s_i,s_{i+1}]$ of a wedge in $\W^i$, then the orientation of the edge $e_{s}$ of $\partial(\ch)$ belongs to the sector formed by the rays $r_i$ and $r_{i+1}$ in~$\os$. Again see Figure~\ref{fig:stabbing}. Also note that if $\os$ contains the supporting lines of the $h$ edges in $\partial(\ch)$, then the stabbing interval of each of the $\W^i$ is a point and therefore $\oh=\ch$.
\end{observation}

It is easy to see that we can calculate the elements $p_{s_1},\dots,p_{s_{2k}}$ on $\partial(\ch)$ in $O(n\log n)$ time; in fact, in $O(k\log n)$ time. This gives us the endpoints of the stabbing interval $[s_i,s_{i+1}]$. Only those intervals containing more than one element are needed; the others can be discarded. Next, we calculate the alternating polygonal chain on $\partial(\oh)$ connecting $p_{s_i}$ to $p_{s_{i+1}}$, which we refer to as a \emph{staircase}.

\subsubsection{Computing the staircases}
\label{subsubsec:staircases-oriented}

\begin{figure}[ht]
  \centering
  {\includegraphics{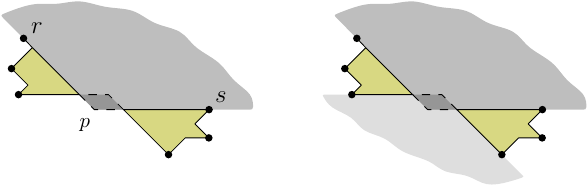}}
  \caption{\label{fig:prelim:rcht}Another $\oht$ for the~$\os$ in Figure~\ref{fig:o-wedges}. Left: An extremal wedge in~$\W^6$ with apex~$p$ and supporting points $r$ and~$s$. Right: Opposite extremal $\os$-wedges, one in~$\W^6$ and the other in~$\W^3$. The overlapping region, which is now a rhomboid instead of a rectangle, is shaded dark gray.}
\end{figure}

The staircase connecting $p_{s_i}$ to $p_{s_{i+1}}$ is determined by the wedges in~$\W^i$ and is contained in the boundary~$\partial(\W^i)$. Counterclockwise around $\oh$, right turns arise at apexes of wedges in~$\W^i$ which we call \emph{extremal}, and left turns arise at points of~$P$ which we call the \emph{supporting points} of those extremal wedges. See Figure~\ref{fig:prelim:rcht}.

Before showing how to compute staircases, let us note that $\oh$ can also be disconnected.
Similarly to Subsection~\ref{sec:urch:oet}, we say that a pair of extremal wedges are \emph{opposite} wedges if one of them is in~$\W^i$ and the other in~$\W^{i+k}$. As shown in~\Cref{fig:prelim:rcht}, a non-empty intersection between two opposite $\os$-wedges results in $\oh$ being disconnected. In this case, we say that the intersecting wedges \emph{overlap}, and refer to their intersection as their \emph{overlapping region}. The following lemma is a straightforward generalization of Lemma~\ref{lemma:overlap_oriented}.

\begin{lemma}\label{lemma:overlap_oriented_o-wedges}
If two extremal $\os$-wedges have a nonempty intersection, then they have to be opposite. When this happens, $\oht$ is not connected.
\end{lemma}

We now proceed with the computation of the staircases, starting by computing their supporting points. In order to do so, we make use of an algorithm by Avis et al.~\cite{theta-maxima_1999}. Let us denote by $\alpha_i$ the angle defined by two consecutive rays $r_i$ and $r_{i+1}$, so  the wedges in~$\W^i$ have angle $\Theta_{i}=\pi-\alpha_i$. We consider $\Theta=\min\{\Theta_i\colon i=1,\dots,2k\}$ and we will distinguish two cases: either $\Theta\geq\frac{\pi}{2}$ or $\Theta<\frac{\pi}{2}$.

Given an angle $\alpha$, a maximal \emph{$\alpha$-escaping wedge} is a wedge of angle $\alpha$ such that (i) its apex is a point $p\in P$, (2) its angle is  at least $\Theta$, and (3) it is $P$-free.

For the first case, when $\Theta\geq\frac{\pi}{2}$, all the sectors defined by the $k$ lines have an angle less than or equal to~$\frac{\pi}{2}$. We use the algorithm by Avis et al.~\cite{theta-maxima_1999} that finds  the set of maximal $\Theta$-escaping wedges in $O(n\log n)$ time and $O(n)$ space.

The points of $P$ that are apexes of a maximal $\Theta$-escaping wedge are called \emph{$\Theta$-maxima}. Indeed, for each such wedge, the algorithm by Avis et al.~\cite{theta-maxima_1999} provides its two defining rays. Thus the algorithm gives at most three maximal \emph{$\Theta$-escaping intervals} for every $p\in P$; hence a \emph{linear number} in total.

\begin{figure}[ht]
\begin{center}
\includegraphics[width=0.8\textwidth]{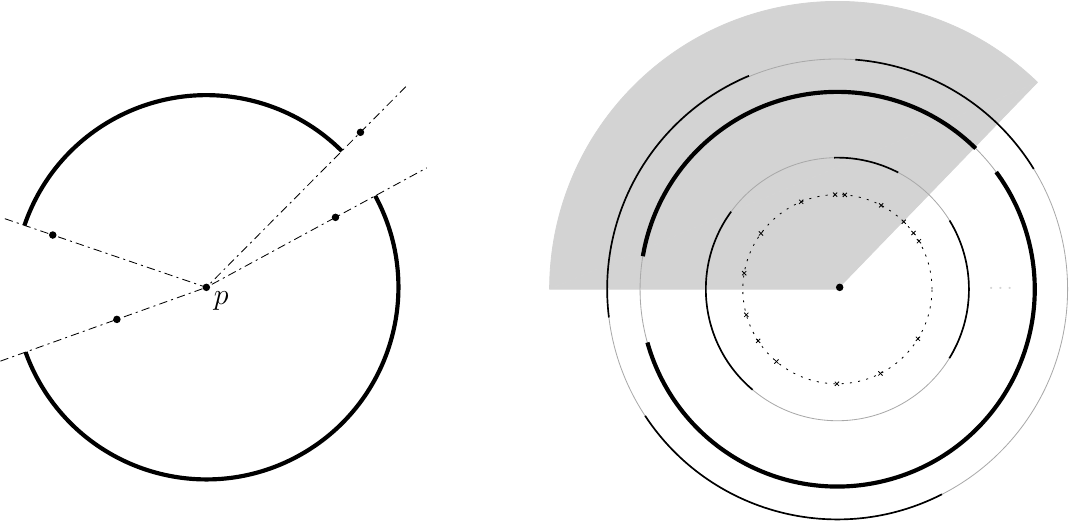}
\end{center}
\caption{\label{fig:dominance_table_sweep} Left: Escaping intervals for a point~$p$ and $\Theta=\frac{\pi}{2}$ as in Figure~\ref{fig:o-wedges}.
Right: Circular table in which the solid circles correspond to the points $p_1,\ldots,p_n$ from inside to outside. On them, the $\Theta$-escaping  intervals; the intervals for the~$p$ depicted on the left are highlighted. In gray, the stabbing interval corresponding to the wedge $W_{{2},{4}}$ from Figure~\ref{fig:o-wedges} (the other stabbing intervals are omitted for the sake of clarity). The innermost circle depicts, as small marks, the vertex events in $[0,2\pi)$ corresponding to the endpoints of the escaping intervals.}
\end{figure}

We will store these intervals in a circular table together with the \emph{stabbing wedges} $W_{{i+1},{i+k}}$. See Figure~\ref{fig:dominance_table_sweep}. In this way, when a stabbing wedge $W_{{i+1},{i+k}}$ fits into the escaping interval of a point $p$, we know that $p$ is not only a $\Theta$-maxima, but actually a $\Theta_i$-maxima, which is indeed equivalent to being a supporting point in $\partial(\W^i)$. (In Figure~\ref{fig:dominance_table_sweep}, right, the gray stabbing interval from the wedge $W_{{2},{4}}$ does not fit into the black escaping interval, because in the left picture the wedge $W_{2}^{4}$ with apex~$p$ cannot escape from~$p$.)

Thus, in $O(n\log n)$ time and $O(n)$ space we can sort the endpoints of the two types of intervals and do a circular sweep of the table, stopping at the defining rays of the wedges~$W_{{i+1},{i+k}}$ to check if the corresponding $p$ supports the staircase contained on~$\partial(\W^i)$. This gives the set $\V$ of vertices of $\oh$.
It remains to obtain the boundary of $\oh$, for which standard techniques~\cite{preparata_1985} can be used in order to compute the staircases $\partial(\W^i)$ from their supporting points and to connect them in $O(n\log n)$ time and $O(n)$ space.
Hence, we have computed $\oh$ in $O(n\log n)$ time and $O(n)$ space.
This time complexity is optimal, since given $\oh$ we can compute $\mathcal{CH}(\mathcal{OH}(P))=\mathcal{CH}(P)$ in linear time, and it is known that computing the convex hull of a set of $n$ points in the plane has an $\Omega(n\log n)$ time lower bound~\cite{preparata_1985}. Therefore, we get the following result.

\begin{theorem}\label{thm:oriented-oh}
Given a set $\os$ of~$k$ lines such that $\Theta\ge\frac{\pi}{2}$, $\oh$ can be computed in optimal $O(n\log n)$ time and $O(n)$ space.
\end{theorem}

When $\Theta<\frac{\pi}{2}$, the algorithm from Avis et al.~\cite{theta-maxima_1999} also works, but it takes $O(\frac{n}{\Theta}\log n)$ time and $O(\frac{n}{\Theta})$ space. The algorithm gives at most $\frac{2\pi}{\Theta}$ maximal \emph{$\Theta$-escaping intervals} for every $p\in P$. Thus we can construct a circular table as above, storing at most $\frac{2\pi}{\Theta}$ circular intervals for each $p$, hence using $O(\frac{n}{\Theta})$ space in total. Therefore, the previous result can be extended as follows

\begin{theorem}\label{thm:oriented-oh-general}
Given a set $\os$ of~$k$ lines such that $\Theta<\frac{\pi}{2}$, $\oh$ can be computed in $O(\frac{n}{\Theta}\log n)$ time and $O(\frac{n}{\Theta})$ space.
\end{theorem}

Note that the value~$\frac{1}{\Theta}$ can be considered a constant for not too small values of~$\Theta$.

\subsection{Computing and maintaining \boldmath{$\oht$}}

Recall that as we rotate $\os$ by an angle~$\theta$ to obtain $\ost$, the wedges $W_{i,j}$ also rotate. Thus, the sets $\mathcal W^i$ change accordingly, giving rise to the sets $\mathcal W^i_{\theta}$. Thus the $\ost$-hull of $P$ is:
\[
\oht =\mathbb{R}^{2} \setminus\bigcup_{i=1}^{2k}\W_\theta^i.
\]

Let $\partial(\W_\theta^i)$ denote the boundary of~$\W_\theta^i$. As in Subsection~\ref{subsubsec:staircases-oriented},
$\partial(\W_\theta^i)$ is an alternating polygonal chain, or staircase, with interior angle $\Theta_{i}=\pi-\alpha_i$. Proceeding counterclockwise along $\oht$, right turns arise at apexes of $\ost$-wedges in $\W_\theta^i$, called extremal, and left turns arise at points of~$P$ which are the supporting points of those extremal wedges.
Recall \Cref{fig:prelim:rcht}.

The following lemma follows directly from Lemma~\ref{lemma:overlap_oriented_o-wedges}.

\begin{lemma}\label{lemma:overlap_unoriented}
If two extremal $\ost$-wedges have nonempty intersection, then they have to be opposite. When this happens, $\oht$ is not connected.
\end{lemma}

As in Subsection~\ref{subsubsec:staircases-oriented}, let $\Theta=\min\{\Theta_i\colon i=1,\ldots,2k\}$ for $\Theta_{i}=\pi-\alpha_i$. We will consider two cases.

The first case is when $\Theta\ge\frac{\pi}{2}$. We show how to maintain $\oht$  for $\theta\in [0,2\pi)$. As in Subsection~\ref{subsec:unorientedRCH}, we will denote by $\St$ the set of overlapping regions in $\oht$, and by $\Vt$ the set of vertices of $\oht$ as $\partial(\W_\theta^i)$ is traversed counterclockwise, $i=1,\dots,2k$.

Applying a rotation of angle $\theta$ to the set $\os$ changes the $\oht$. In particular, the supporting vertices of the staircases $\partial(\W_\theta^i)$ might change. We now show how to update these staircases in $O(\log n)$ time by insertion or deletion of a point. In order to do so, we need to maintain the (at most) $2k$ staircases in (at most) $2k$ different balanced trees, one for each staircase. Note that some of the staircases may appear and/or disappear during the rotation. The total insertion or deletion operations can be done in $O(kn\log (kn))=O(kn\log n)$ time.

Referring to the circular table in Figure~\ref{fig:dominance_table_sweep}, we can rotate the (gray) stabbing wedges~$W_{{i+1},{i+k}}$, stopping at events arising when a defining ray of a stabbing wedge hits a vertex event in the innermost circle; i.e., entering or leaving an escaping interval (black). This provides the information about whether the stabbing wedges fit into the escaping intervals or not, and this, as in Subsection~\ref{subsubsec:staircases-oriented}, enables us to handle the insertion or deletion of points in the set~$\Vt$ of vertices of $\oht$ (i.e., the points on the staircases). Since the number of escaping intervals for a point is at most three and during the rotation these can arise in any of the $2k$ wedges corresponding to rotated $W_{{i+1},{i+k}}$, there are $O(kn)$ events. Thus we obtain the following result.

\begin{theorem}\label{thm:unoriented-oh}
Given a set $\os$ of~$k$ lines such that $\Theta\ge\frac{\pi}{2}$, computing and maintaining the boundary of $\oht$ during a complete rotation of $\theta\in [0,2\pi)$ can be done in $O(kn\log n)$ time and $O(kn)$ space.
\end{theorem}

Furthermore, it is straightforward to see that we can keep track of the parameters of $\oht$ as the algorithm runs.

\begin{corollary}\label{coro:steps-components_1}
Given a set $\os$ of~$k$ lines such that $\Theta\ge\frac{\pi}{2}$, computing an orientation $\theta$ such that the boundary of $\oht$ has the minimum number of steps, has the minimum number of staircases,  is connected, or has the minimum number of connected components, can be done in $O(kn\log n)$ time and $O(kn)$ space.
\end{corollary}

It is easy to see, as we did in Section~\ref{subsec:oriented_Ohull}, that when $\Theta<\frac{\pi}{2}$, the next results follow:

\begin{theorem}\label{thm:unoriented-oh-general}
Given a set $\os$ of~$k$ lines such that $\Theta<\frac{\pi}{2}$, computing and maintaining the boundary of $\oht$ during a complete rotation of $\theta\in [0,2\pi)$ can be done in $O(k\frac{n}{\Theta}\log n)$ time and $O(k\frac{n}{\Theta})$ space.
\end{theorem}

\begin{corollary}\label{coro:steps-components_2}
Given a set $\os$ of~$k$ lines such that $\Theta<\frac{\pi}{2}$, computing an orientation $\theta$ such that the boundary of $\oht$ has the minimum number of steps, has the minimum number of staircases, is connected, or has the minimum number of connected components, can be done in $O(k\frac{n}{\Theta}\log n)$ time and $O(k\frac{n}{\Theta})$ space.
\end{corollary}

Another interesting consequence of Theorem~\ref{thm:unoriented-oh-general} is that we can also generalize Theorem~\ref{mantenimiento} to the case where~$k=2$ and the two lines are not perpendicular.

\begin{corollary}\label{coro:2directions}
For a set $\os$ containing only two non-perpendicular lines, computing and maintaining the boundary of $\oht$ during a complete rotation for $\theta\in [0,2\pi)$ can be done in $O(\frac{n}{\Theta}\log n)$ time and $O(\frac{n}{\Theta})$ space, where $\Theta$ is the smallest angle of the sectors defined by the two lines.
\end{corollary}

\subsection{Finding the value of \boldmath{$\theta$} for which \boldmath{$\oht$} has minimum area}\label{subsec:area_oh}

The results in Subsections~\ref{subsec:unorientedRCH} and~\ref{sec:area} can be adapted to the case of a set $\os$ of $k$ lines through the origin such that all the sectors they define have angle at most~$\frac{\pi}{2}$; i.e., with $\Theta\ge\frac{\pi}{2}$. Again, for a fixed value of~$\theta$, we can compute the area of $\oht$ using the fact that
\begin{equation}
  \label{eq:prelim:area_oh}
  \area(\oht) = \area(\polygon) - \area(\polygon \setminus \oht).
\end{equation}

As before, we will compute the area of $\polygon\setminus\oht$ by decomposing it into two types of regions: (i)~the triangles defined by every pair of consecutive elements in $\Vt$, and (ii)~the rhomboid overlaps between two triangles which make $\oht$ disconnected. Recall Figures~\ref{fig:prelim:area} and~\ref{fig:prelim:rcht}.

By Theorem~\ref{thm:unoriented-oh}, the triangles in~(i) above can be maintained in $O(kn\log n)$ time and $O(kn)$ space. As $\theta$ increases from~$0$ to~$2\pi$, the set $\Vt$ of points on~$\partial(\oht)$ changes at the values of~$\theta$ where a point of $P$ becomes (resp.\ is no longer) a vertex of $\oht$. These angles are again called \emph{insertion} (resp.\ \emph{deletion}) \emph{events}.

Next, we will deal with the rhomboids in~(ii), and show how to maintain the set $\St$ of rhomboid overlaps which, analogously to~$\Vt$, changes at \emph{overlap} (resp.\ \emph{release}) \emph{events}. We will use the same techniques as in Subsection~\ref{subsec:unorientedRCH}, but repeat the process at most $k$ times; i.e., separately computing all the possible overlapping rhomboids for each pair of opposite staircases in~$\oht$ as $\theta$ increases from $0$ to $2\pi$.

\subsubsection{The sequence of overlap and release events}\label{sec:oh:oet}

Given an angle $\theta$, let us denote an overlap in $\oht$ as $S_{\theta}(i,j)$, and let $\mathcal{S}_\theta(P)$ be the set of all overlaps $S_{\theta}(i,j)$ of $\oht$. By Theorem~\ref{thm:unoriented-oh}, for any $\theta$, $\oht$ can be computed in $O(kn\log n)$ time and $O(kn)$ space, and $\mathcal{S}_\theta(P)$ can then be computed from $\Vt$ in $O(n)$ time.

To maintain the set $\mathcal{S}_{\theta}(P)$ while $\theta$ increases from~$0$ to~$2\pi$, we will compute a sequence of events for each pair of opposite staircases, computing their overlap and release events. The events of each pair are computed independently, and then the $k$ sequences are merged in $O(kn)$ time to obtain the set of all overlap and release events.

By Lemma~\ref{lemma:overlap_unoriented}, only opposite staircases $\partial(\W_\theta^i)$ and $\partial(\W_\theta^{i+k})$ can intersect, making $\oht$ disconnected. The corresponding intersection can be composed of several overlapping regions, which we will now show how to maintain as $\theta$ increases.

We will adapt the definitions and observations in Subsection~\ref{subsubsec:chain_arcs}, highlighting the differences. First, the chain of arcs is defined in essentially the same way, the only difference being that since the loci of the points are apexes of $P$-free extremal $\theta$-wedges, they might be ``flatter arcs,'' with curvature smaller than or equal to that of circle arcs, since the angles of the wedges are now at least~$\frac{\pi}{2}$ instead of exactly~$\frac{\pi}{2}$. That the arcs may be flatter does not affect the monotonicity of any sub-chain~$A_{e_i}$ with respect to an edge~$e_i$ of~$\partial(\ch)$. The property of overlapping regions corresponding to intersecting links is also maintained. The reader may refer to Figure~\ref{fig:cor:arc-chain}, considering the possibility that the arcs may be flatter.

Next, we adapt Subsection~\ref{subsection:linear} to see that the number of intersections between links is now in~$O(kn)$. For this, it is enough to check that Lemmas~\ref{lem:linear:angle},~\ref{lem:linear:link_inter}, and~\ref{lem:linear:link_inter_linear} are valid for each of the $k$ pairs of opposite staircases in~$\oht$ and,
using them, Theorem~\ref{thm:linear:linear} is still true for each of the~$O(k)$ pairs of opposite staircases in~$\oht$.
It is important to notice that the assumption of $\Theta\ge\frac{\pi}{2}$ in Subsection~\ref{subsec:area_oh} is needed for the results in the last two paragraphs to work.

Now we adapt Subsection~\ref{subsection:table} to compute the sequence of overlap and release events. In order to do so, we proceed with an algorithm analogous to the one outlined there, but computing the sequence of overlap and release events for each pair of opposite staircases in $O(n\log n)$ time and $O(n)$ space. After doing so, we merge the $k$ sequences of overlap and release events obtained into a single sequence of $O(kn)$ overlap and release events, obtaining the sorted events during a complete rotation of~$\theta$ from~$0$ to~$2\pi$. In this way, we get the following result, which generalizes Theorem~\ref{thm:table:overlap}.

\begin{theorem}\label{oht:thm:table:overlap}
The sequence of $O(kn)$ overlap and release events for~$\oht$ as~$\theta$ increases from~$0$ to~$2\pi$ can be computed in $O(kn\log n)$ time and $O(kn)$ space.
\end{theorem}

In order to sweep the sequence of overlap and release events, we again store the event sequence as points on a circle, on which we represent the wedges $W_{{i+1},{i+k}}$ in way similar to the way we did in the innermost circle in Figure~\ref{fig:dominance_table_sweep}, right (where  we stored the vertex events instead). Proceeding as in Subsection~\ref{subsection:maintain}, but now considering that we have at most $O(kn)$ total overlap and release events, we have the following result, which generalizes Theorem~\ref{thm:maintain:maintain}.

\begin{theorem}\label{oht:thm:maintain:maintain}
Using the sequence of overlap and release events for $\oht$ as $\theta$ increases in $[0,2\pi)$, the set $\St$ can be maintained  using $O(kn)$ time and $O(kn)$ space.
\end{theorem}

\subsubsection{Computing minimum area}\label{sec:area_oh}

The final step is to compute the value of $\theta$ that minimizes (or maximizes) the area of $\oht$. We next show how to compute this angle in $O(kn\log n)$ time and $O(kn)$ space.

Let $\alpha$, $\beta$ be two events (of any type) such that $(\alpha,\beta)$ is an angular interval in $[0,2\pi)$ containing no
events. Extending Equation~\ref{eq:prelim:area_oh} and mimicking Equation~\ref{eqn:area:area}, we express the area of $\oht$ for any $\theta\in (\alpha,\beta)$ as
\begin{equation}
  \label{eqn:area:area_oh}
  \area(\oht)
  \ =\ \area(\polygon)
  \ -\ \sum_{i} \area(\triangles)
  \ +\ \sum_{j} \area(\lozenge_j(\theta)).
\end{equation}

\begin{figure}[ht]
  \centering
  \subcaptionbox{\label{fig:triangle:1}}
  {\includegraphics{o-wedge-4}\vspace{1.5em}}
  \qquad{}
  \subcaptionbox{\label{fig:triangle:2}}{
    \includegraphics{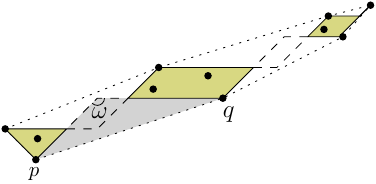}}
  \caption{Computing the area of a triangular region.~\subref{fig:triangle:1} The wedge of size $\omega$ that bounds the triangle.~\subref{fig:triangle:2} The triangle defined by $p$ and $q$.}\label{fig:triangle}
\end{figure}

\begin{figure}[ht]
  \centering
  \subcaptionbox{\label{fig:overlapping-region:1}}
  {\includegraphics{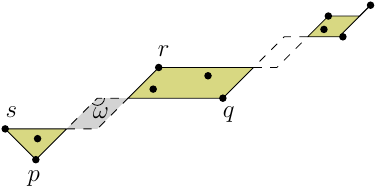}}
  \quad{}
  \subcaptionbox{\label{fig:overlapping-region:2}}
  {\includegraphics[scale=0.9]{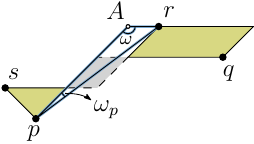}}
  \quad{}
  \subcaptionbox{\label{fig:overlapping-region:3}}
  {\includegraphics[scale=0.9]{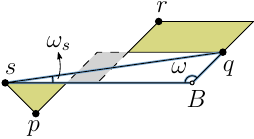}}
  \caption{Computing the area of an overlapping region.~\subref{fig:overlapping-region:1} The overlapping region defined by $p,q,r$, and $s$.~\subref{fig:overlapping-region:2} The triangle $\triangle prA$.~\subref{fig:overlapping-region:3} The triangle $\triangle qsB$.}\label{fig:overlapping-region}
\end{figure}

The term $\triangles$ denotes a triangle defined by two consecutive vertices $p,q \in \Vt$. The boundary of this triangle is formed by the line segment $\overline{p q}$ and one of the current $\ost$-wedges in $\W_\theta^i$ supported by $p$ and $q$. As the angle of an extremal wedge is at least $\frac{\pi}{2}$, the triangle is now either rectangular
(Figure~\ref{fig:prelim:area}) or obtuse (Figure~\ref{fig:triangle}). Finally, the term $\lozenge_j(\theta)$
denotes the $j$-th overlapping region in $\St$, which is now a rhomboid (Figure~\ref{fig:overlapping-region}).

We now show that for any particular value of $\theta$, we can evaluate \Cref{eqn:area:area_oh} in $O(kn\log n)$ time, and as
$\theta$ increases from $0$ to $2\pi$, a constant number of terms need to be updated at each event point, regardless of its type.

\paragraph{The polygonal region.}

At any fixed value of~$\theta$ the area of $\polygon$ can be computed from $\Vt$ in $O(n)$ time. The term $\area(\polygon)$ changes only at vertex events. These events can be processed in constant time, since at an insertion (resp.\ deletion) event, the area of a single triangle needs to be subtracted (resp.\ added) to the previous value of $\area(\polygon)$.

\paragraph{The triangles.}

Let $p$ and $q$ be two consecutive vertices in $\Vt$ such that $p$ precedes $q$.~Suppose that for any $\theta \in(\alpha,\beta)$, the points $p$ and $q$ define the triangle $\triangle_i(\theta)$. This triangle is bounded by $\overline{p q}$ and an extremal wedge supported by $p$ and $q$. Let $\omega \geq \frac{\pi}{2}$ denote the angle of the extremal wedge, and $\omega_p$ and $\omega_q$ denote respectively the internal angles of $\triangle_i(\alpha)$ at $p$ and $q$. See Figure~\ref{fig:triangle}.

The area of $\triangle_i(\theta)$ can be expressed as
\begin{equation}
  \label{eq:triangle_1}
  \area(\triangle_i(\theta)) = \frac{\vert \overline{pq} \vert^2}{2 \sin(\omega)} \cdot \sin(\omega_p +
  (\theta - \alpha)) \cdot \sin(\omega_q-(\theta-\alpha)),
\end{equation}
and expanding \Cref{eq:triangle_1} we obtain
\begin{align}
  \label{eq:triangle_2}
  \area(\triangle_i(\theta))
  &= A \cdot \cos^2(\theta - \alpha) + B \cdot \cos(\theta - \alpha) \cdot \sin(\theta - \alpha) + C \cdot \sin^2(\theta - \alpha)\nonumber\\
  &= D + E \cdot \cos(2(\theta - \alpha)) + F \cdot \sin(2(\theta - \alpha)),
\end{align}
where $A \ldots F$ are constant values in terms of $\omega$, $\omega_p$, $\omega_q$, and the coordinates of $p$ and $q$. Therefore, the term $\sum_i \area(\triangle_i(\theta))$ is linear in $\cos(2(\theta - \alpha))$ and $\sin(2(\theta - \alpha))$. Since each point of $P$ can appear in $O(k)$ staircases, we have $O(kn)$ triangles, which can be processed in $O(kn)$ time as we described in Subsection~\ref{subsec:unorientedRCH}

\paragraph{The overlapping regions.}

Let $p,q$ be two consecutive vertices in $\Vt$ such that $p$ precedes $q$, and let $r,s$ be two consecutive vertices in $\Vt$ such that $r$ precedes $s$.~Suppose that for any $\theta \in (\alpha,\beta)$, the points $p,q,r,s$ define the overlapping region
$\lozenge_j(\theta)$. Without loss of generality, we assume that $p$ and $q$ support an extremal wedge in $\mathcal{W}^i$, and that $r$ and $s$ support an extremal wedge in $\mathcal{W}^{i+k}$. We denote by $\omega \geq \frac{\pi}{2}$ the angle of both wedges. See Figure~\ref{fig:overlapping-region}\subref{fig:overlapping-region:1}.

Let us consider the overlapping region when $\theta=\alpha$. Based on the triangles $\triangle prA$
(Figure~\ref{fig:overlapping-region}\subref{fig:overlapping-region:2}) and $\triangle qsB$
(Figure~\ref{fig:overlapping-region}\subref{fig:overlapping-region:3}), the area of $\lozenge_j(\theta)$ can be expressed as
\begin{equation}
  \label{eq:overlaps_1}
  \lozenge_j(\theta) =
  \frac{\vert \overline{p r} \vert \cdot \vert \overline{q s} \vert}{\sin^2(w)} \cdot
  \sin (\omega_p + (\theta - \alpha)) \cdot
  \sin (\omega_s - (\theta - \alpha)),
\end{equation}
and expanding Equation~\ref{eq:overlaps_1} we obtain
\begin{align}
  \label{eq:overlaps_2}
  \area(\triangle_i(\theta))
  &= A^{\prime} \cdot \cos^2(\theta - \alpha) + B^{\prime} \cdot \cos(\theta - \alpha) \cdot \sin(\theta - \alpha) + C^{\prime} \cdot \sin^2(\theta - \alpha)\nonumber\\
  &= D^{\prime} + E^{\prime} \cdot \cos(2(\theta - \alpha)) + F^{\prime} \cdot \sin(2(\theta - \alpha)),
\end{align}
where $A^{\prime} \ldots F^{\prime}$ are constant values in terms of $\omega, \omega_p, \omega_s$ and the coordinates of $p$, $q$, $r$, and $s$. The term $\sum_i \area(\lozenge_i(\varphi))$ is therefore linear in $\cos(2(\theta - \alpha))$ and $\sin(2(\theta - \alpha))$. We thus can have at most $O(kn)$ overlapping regions in the sequence of overlap and release events, and they can be processed in $O(kn)$ time as in Subsection~\ref{sec:area}. Notice that the area of a rhomboid can be computed in constant time. Using the algorithm in Subsection~\ref{subsubsec:search_algorithm} and taking into consideration the changes just mentioned, we get our final result.

\begin{theorem}\label{teorem_4_1}
Given a set $\os$ of~$k$ lines such that $\Theta\ge\frac{\pi}{2}$, computing $\oht$ with minimum (or maximum) area over all
$\theta\in [0,2\pi)$ can be done in $O(kn\log n)$ time and $O(kn)$ space.
\end{theorem}

\end{document}